\documentclass[acmsmall]{acmart}

\usepackage{booktabs}
\usepackage{pifont}
\usepackage{placeins} 
\usepackage{tabularx}
\usepackage{microtype}
\usepackage[normalem]{ulem}
\usepackage{siunitx} 
\usepackage{graphicx}
\usepackage{array}        
\usepackage{colortbl}     
\usepackage{paracol}
\usepackage{xcolor}       
\usepackage{balance}
\RequirePackage[l2tabu, orthodox]{nag}
\usepackage{microtype}
\usepackage{lipsum} 

\usepackage{subcaption}
\usepackage{tikz} 
\usetikzlibrary{shapes.geometric, arrows}
\usepackage{multirow}
\usepackage[subtle]{savetrees} 
\usepackage{caption}
\usepackage{enumitem}
\usepackage{color,soul}
\usepackage[normalem]{ulem}
\newcommand\VRule[1][\arrayrulewidth]{\vrule width #1}
\newcommand{\studyquote}[2]{{\def\arraystretch{2}\setlength\tabcolsep{7pt}\vspace{0.5ex} \noindent\begin{tabular}{!{\color{gray!75}\VRule[2pt]}p{\dimexpr\linewidth-2\tabcolsep-0.3pt}}\cellcolor{gray!10}\textit{``#1''} \mbox{{(P#2)}}\tabularnewline\end{tabular}\vspace{1ex}}}

\author{Zainab Iftikhar}
\orcid{0000-0002-4086-436X}
\affiliation{%
  \institution{Brown University}
  \department{Department of Computer Science}
  \city{Providence}
  \country{USA}
}

\author{Sean Ransom}
\orcid{0009-0009-0135-244X}
\affiliation{%
  \institution{Louisiana State University Health Sciences}
  \department{Department of Psychiatry}
  \city{New Orleans}
  \country{USA}
}

\author{Amy Wei Xiao}
\orcid{0009-0003-2303-8204}
\authornote{Work began while author was at Brown University}
\affiliation{%
  \institution{University of Washington}
  \department{Human Centered Design \& Engineering}
  \city{Seattle}
  \country{USA}
}

\author{Nicole Nugent}
\orcid{0000-0001-8756-5618}
\affiliation{%
  \institution{Alpert Medical School of Brown University}
  \department{Department of Psychiatry and Human Behavior}
  \city{Providence}
  \country{USA}
}

\author{Jeff Huang}
\orcid{0000-0002-3453-5666}
\affiliation{%
  \institution{Brown University}
  \department{Department of Computer Science}
  \country{USA}
}

\renewcommand{\shortauthors}{Zainab Iftikhar, Sean Ransom, AmyWei Xiao, Nicole Nugent, \& Jeff Huang}

\newcommand{\cscwmajorrevision}[1]{{\textcolor{black}{{#1}}}}
\newcommand{\cscwfinal}[1]{{\textcolor{black}{{#1}}}}

\setcopyright{cc}
\setcctype{by}
\acmJournal{PACMHCI}
\acmYear{2026} \acmVolume{10} \acmNumber{6} \acmArticle{CSCW196}
\acmMonth{10} \acmDOI{10.1145/3817112}

\graphicspath{{figures/}}

\begin{document}
\title[Therapy as an NLP Task: Comparing LLMs and Human Peers Behaviors in CBT Sessions]{Therapy as an NLP Task: Comparing LLMs and Human Peers Behaviors in CBT Sessions}

\begin{abstract}
Large language models (LLMs) are increasingly being used as ad hoc therapists. While prior research has found that LLMs outperform human counselors in generating single-turn empathetic responses, fewer studies have compared their behaviors across multi-turn sessions. In this study, we compare the session-level behaviors of human peer counselors with those of an LLM, both trained on the same manual to deliver multi-turn, single-session Cognitive Behavioral Therapy (CBT). Our three-phase, mixed-methods study involved: (a) an 18-month ethnography of a peer support platform, where seven counselors iteratively refined CBT prompts through 110 self-counseling sessions and 60 weekly focus groups; (b) a novel session generation method that allows direct, controlled comparison of human and LLM counselors under matched conditions---client responses were drawn from publicly available human-led CBT sessions while counselor responses were generated by a CBT-prompted LLM; and (c) expert evaluations conducted by three licensed clinical psychologists. Through data triangulation, our results show a trade-off. Human peer counselors use relational techniques to interpret subtle cues, adapt CBT to users' values and cultural contexts, and use strategies such as small talk and contextually relevant self-disclosure to build rapport and guide the session, but often at the expense of session structure and therapeutic focus. LLM counselors, on the other hand, demonstrate greater methodological adherence to CBT techniques, but struggle to sustain turn-taking, frequently fail to distinguish between clinically important and trivial content, and are more prone to lecturing and imposing solutions. LLM counselors also tend to produce ``deceptive empathy'', excessively anthropomorphic responses that can inflate user expectations of genuine human care. Taken together, our findings imply that while LLMs may outperform human counselors when generating a single-turn interaction, their ability to lead multi-turn sessions is more limited, highlighting that therapy cannot be reduced to a standalone natural language processing (NLP) task. We conclude by mapping concrete design opportunities and ethical guardrails for hybrid human-AI systems, emphasizing the risks of over-attributing human relational subjectivity to current LLMs.
\end{abstract}


\maketitle
\section{Introduction} 

Due to systemic barriers, many individuals seeking therapy have turned to large language models (LLMs) for support and guidance~\cite{ma2024evaluating, aktan2022attitudes, heinz2025randomized}, even though these models are largely unsanctioned. Likewise, organizations like the National Eating Disorders Association even replaced their helpline staff with a wellness chatbot, only to suspend it five days later. Framed as tools for efficiency and scalability, the promise and potential of LLMs for mental health support have prompted a growing interest in comparing them with human mental health counselors~\cite{morris2018towards, aktan2022attitudes, young2024role, vowels2024chatbots, hatch2025eliza}. 

Much of the research comparing humans with LLMs has focused on their ability to respond to a user in an isolated one-time exchange (single-turn interaction), showing that the content generated by an LLM is not only rated higher on psychotherapy principles~\cite{hatch2025eliza}, but is also preferred over messages written by human experts~\cite{morris2018towards, aktan2022attitudes, ayers2023comparing, young2024role, vowels2024chatbots}. However, such experiments compare user preferences toward a single empathetic response to a user post~\cite{young2024role, vowels2024chatbots, ayers2023comparing, morris2018towards}, or use vignettes involving fictional characters in therapy~\cite{aleem2024towards, hatch2025eliza}, overlooking the complete \textit{context} of a user's lived experience and the longitudinal relationship within the therapeutic session.

The isolated nature of single-turn evaluations makes it difficult to generalize the findings to the model's behavior across an entire session. Psychotherapy, or talk therapy, is not just a single-turn interaction (a one-time response to a user's post), but a continuous, multi-turn dialogue between a counselor and the client. Users seeking talk therapy are unlikely to receive a response to their message and stop at that; in fact, there have been instances where prolonged, multi-turn interactions with LLMs have led to tragic outcomes~\cite{xiang2023eating, gerken2024chatbot, cole2023queen}. 

Without session-level comparisons, we cannot observe LLMs' back-and-forth conversational dynamics throughout a session, nor can we answer what a hybrid model of human-AI collaboration looks like, a concept frequently proposed in the digital mental health literature but rarely specified at the session level~\cite{sharma2023human, ayers2023comparing, syed2024machine, young2024role, heinz2025randomized}. While there are many ``units'' in talk therapy---such as individual responses, entire sessions (synchronous back-and-forth discourse with many statements and responses), and multiple sessions for a client (with regular intervals over time)---this paper focuses on single sessions in the context of CBT, which is a psychotherapeutic treatment designed to help clients reframe negative thinking patterns. Single-session interventions are one-off, goal-oriented and structured sessions, often lasting an hour, during which a counselor provides brief 1:1 support to address specific issues, rather than requiring multiple visits~\cite{dryden2022single, ellis2015treating, fitzpatrick2017delivering, syed2024machine}. 

Given the scale at which LLMs are being prompted to act as therapists (a scenario we refer to as an LLM counselor), we compare the behavior of such models across a session (from start to finish) with that of human counselors and examine the implications for users' therapeutic experience, asking the following research questions:

\begin{itemize}
    \item RQ1: How do human peer counselors and LLM counselors, both trained in CBT principles, compare in delivering effective single-session interventions?

    \item RQ2: What specific and significant challenges are faced by each, according to clinical psychologists trained in evaluating CBT sessions?

    \item RQ3: What design lessons can we learn from comparing human and LLM counselors' session-level behaviors, and how can these insights help us reflect on human-AI collaboration for scaling mental health support?
\end{itemize}

To answer these questions, we conducted a mixed-methods study
with seven trained peer counselors and three licensed clinical psychologists. Our study was designed in three phases: (1) an 18 month long naturalistic ethnographic study based on a peer support platform, where peer counselors iteratively tested and refined system prompts for several LLM architectures through 110 self-counseling sessions and 60 weekly focus groups, (2) a novel synthetic session generation method enabling direct controlled comparison of human and LLM counselors under matched conditions, where client responses are drawn from transcripts of 27 publicly available human-led CBT sessions to generate 27 LLM-counseled sessions to avoid patient harm, and (3) session evaluations conducted by three licensed clinical psychologists trained in implementing and evaluating CBT sessions. Sessions are experimentally generated instead of re-conducted to avoid a setup that has been documented to raise serious ethical concerns~\cite{biron2023online} and can cause harm to vulnerable populations. For example, in their randomized controlled trial,~\citet{heinz2025randomized} asked clinicians to manually review all messages sent by the ``therapy'' chatbot to correct harmful responses, questioning the reliability of fully autonomous LLM counselors.

As we will show through session examples and 
data triangulation, human and LLM counselors bring contradictory strengths and limitations to single-session interventions. Human counselors use relational strategies to guide the session, are able to interpret subtle cues that lead to a greater understanding of a user's lived experience, and adapt CBT methods to align with users' values (RQ1). However, these sessions often include non-therapeutic elements (excessive small talk and inappropriate self-disclosure) that detract from the session's therapeutic focus (RQ2). In contrast, LLM-led sessions tend to offer more psychoeducation and greater adherence to the CBT method, but often at the expense of turn-taking, cultural understanding and flexibility (RQ1). LLM counselors also struggle to distinguish clinically important user disclosures from less relevant information (RQ2).

While recent studies suggest that LLM responses exhibit greater empathy and quality than those written by humans due to their longer length and positive sentiment~\cite{morris2018towards,ayers2023comparing,young2024role, hatch2025eliza}, our analysis suggests the opposite: longer responses can undermine therapeutic collaboration with LLM counselors dominating the conversation, creating a power imbalance that can negatively affect psychotherapeutic elements such as therapeutic alliance and CBT application (RQ2). Combining these findings, we discuss how a hybrid model of care can combine human interpersonal effectiveness with LLM's adherence to method to scale therapeutic support (RQ3). Lastly, we found that alternative therapeutic methods, whether provided by human peers or LLM counselors, differ from traditional CBT administered by licensed clinicians, as both at times deviate from standard therapeutic protocols in ways that may pose a risk of patient harm (RQ3).
\newline

\textbf{Note:} While we use the term ``therapy'' in the title to frame the research question under examination, in discussing sessions we deliberately avoid such terms, including `talk therapy' and ``psychotherapy''. Unlike prior work that refers to LLM-based mental health support as talk therapy or LLM counselors as therapists, we argue that the support provided by an LLM counselor, even when prompted to follow therapeutic techniques, does not constitute therapy which is a clinical practice with legal licensing.


\section{Background \& Related Work}
Alternative cost-effective interventions, including peer counseling platforms~\cite{wang2012stay, o2016turning, kornfield2022involving, smit2023effectiveness, syed2024machine} and AI-based CBT counseling~\cite{fitzpatrick2017delivering, morris2018towards, inkster2018empathy, gaffney2019conversational, prochaska2021therapeutic}, have emerged within the CSCW literature as promising avenues to increase user access to mental health support. This section outlines some of the prior and current work in the field.

\subsection{Peer-based Counseling for Mental Health Support}
The United States has an average of thirty psychologists per one hundred thousand people~\cite{who2019report}. This ratio is unlikely to increase through training additional professionals alone. Instead, new scalable approaches are emerging to expand access, including peer counseling platforms~\cite{o2016turning, kornfield2022involving}. \textit{Peer counseling}, also known as peer support or online peer-to-peer therapy~\cite{yao2022learning}, is defined as:
\begin{quote}
    A system of giving and receiving help founded on key principles of respect, shared responsibility, and mutual agreement on what is helpful~\cite{mead2006peer}
\end{quote}

In the US, peer counseling groups, self-help organizations, and consumer-operated services are more than double the traditional and professional mental health organizations~\cite{goldstrom2006national}. Initially, groups such as Alcoholics Anonymous (AA)~\cite{yarosh2013shifting}, InTheRooms.com (ITR)~\cite{rubya2017video}, GROW and eGrow~\cite{young1987evaluation} started as community-based organizations claiming that individuals with similar lived experiences can better relate to each other and offer more genuine understanding, empathy, and validation~\cite{ smit2023effectiveness, kornfield2022involving}. Peer counseling has also been shown to be effective for clinical and personal recovery across a wide range of mental disorders and intervention types~\cite{smit2023effectiveness}. This led to the shaping of current peer support counseling through digital innovations, including crowdsourcing empathetic responses to a user's posts~\cite{morris2018towards, kornfield2022involving} or research into unmoderated online communities and social networks for support seeking~\cite{naslund2016future}. For instance, \citet{morris2015efficacy} developed Koko, a message-based peer support system, to crowd-source one-off interactions for users' help-seeking messages. The platform's design was inspired by Panoply, a web-based peer support platform that was demonstrated to alleviate depressive symptoms. Other researchers have focused on how context-specific self-disclosure in online communities also helps build support~\cite{wang2012stay, andalibi2017sensitive}. 

While peer counseling holds potential to shape the future of mental health support~\cite{ naslund2016future, smit2023effectiveness}, counselors often lack formal training~\cite{yao2022learning, morris2015efficacy}. Unlike trained professionals who receive extensive psychotherapeutic training, peers connect with individuals through shared lived experiences~\cite{mead2006peer}. This support can foster a sense of understanding and connection, but the interactions can suffer from a lack of evidence-based content.

Hence, there is an increasing emphasis within CSCW on training peers via manuals on online platforms~\cite{morris2015efficacy, o2017design, rubya2017video, smith2021effective, yao2022learning}. For example, platforms like 7 Cups of Tea and Koko offer training courses for CBT techniques like active listening; whereas the Cheeseburger Therapy platform requires 
its peers to undergo CBT training, focusing on active listening, reflective restatements, and cognitive reappraisal~\cite{syed2024machine} and then uses cohort socialization processes in which CBT trained members onboard a new peer counselor in roleplay-based practice counseling sessions. Training peers for single-session interventions has been found to be effective in existing research. For instance,~\citet{syed2024machine} found that CBT training helps peers provide higher empathetic support. Though effective, this reliance on 1:1 human counselors circles back to the ongoing issue of access: the limited availability of trained people to meet treatment needs, leading researchers to look at AI-based alternatives~\cite{ly2017fully, fitzpatrick2017delivering, prochaska2021therapeutic}.

\subsection{Therapy Bots: From Rule-based Agents to Large Language Models}
In response to the limited availability of trained counselors, prior work has increasingly focused on using AI and NLP techniques for talk-based therapy, particularly those based on CBT principles. A significant area of focus is the use of conversational agents or chatbots for standalone support~\cite{ly2017fully, gaffney2019conversational, ebert2019digital, chekroud2021promise}, either as agents providing psychoeducational content or as a psychotherapist~\cite{prochaska2021therapeutic, stade2024large, heinz2025randomized}. 

Woebot and Wysa are two well-known chatbots that use CBT principles and were developed with input from mental health professionals. Woebot, a text-based conversational agent, delivers CBT content in a conversational format and has been used to self-manage depressive symptoms~\cite{fitzpatrick2017delivering} and substance use disorders~\cite{prochaska2021therapeutic}. Similarly, users of Wysa, an AI-based mental health chatbot, reported lower levels of depressive symptoms after engaging with the chatbot~\cite{inkster2018empathy}. Other tools include standalone CBT interventions (that take the role of a psychotherapist) to improve well-being for non-clinical population~\cite{morris2018towards, smith2021effective, chekroud2021promise}. 

However, such agents were rule-based, meaning they followed predefined scripts, limiting their ability to adapt to dynamic human behavior and tailor responses to individual needs~\cite{laranjo2018conversational, graham2019artificial, fiske2019ethical}. The lack of personalization and natural language capability presented a challenge to their uptake, as psychotherapy is client-centric and relies heavily on personalized conversations for effective treatment. And so, LLMs, due to their personalized conversational fluidity and high engagement, gained significant attention. 

Though LLMs lack genuine understanding, they are highly effective at generating tailored responses to user inputs in a near-conversational style~\cite{gu2023mentalblend, na2024cbt, ji2022mentalbert}. The ability to generate human-like language, combined with user-friendly interfaces, helped overcome the engagement issues of traditional rule-based chatbots, allowing thousands of users to customize publicly available LLM-based chatbots to address their mental health needs without professional guidance. Given this growing interest, researchers have focused on the potential of LLMs within mental health contexts, ranging from prompt design to standalone CBT intervention~\cite{ vowels2024chatbots, ma2024evaluating, heinz2025randomized, kim2024mindfuldiary, na2024cbt}. In their randomized controlled clinical trial,~\citet{heinz2025randomized} found that the LLM-based chatbot resulted in clinically significant improvements in participants' mood and overall well-being for users with depression and anxiety, suggesting its potential to act as an independent solution in psychotherapy.

\subsection{Technical and Ethical Challenges in Repurposing LLMs for Mental Health Counseling}
Repurposing LLMs for mental health support when they were originally intended for basic language generation presents several challenges. First, LLMs are designed to predict the next token in a sequence from a given text based on patterns learned during training, which are largely devoid of fact-checking~\cite{brown2020language}. This encompasses traditional challenges of AI-mediated healthcare, including a lack of high-quality training data, low external validity, misinformation, and societal biases~\cite{ernala2019methodological}. In addition, there are specific challenges to deploying LLMs in clinical settings, including model hallucination, toxic content, and racial and gender bias~\cite{xu2024mental, pentina2023exploring, ma2024understanding, si2022so}. Despite recognizing various benefits, such as emotional support and reducing practitioners' workload, these inherent complexities raise stakeholder concerns~\cite{stade2024large, jo2023understanding}.

Moreover, there are ethical and epistemic challenges around these systems' capacity to provide safe emotional support. Recent research in bioethics has particularly highlighted the harmful ethical implications of conversational agents~\cite{martinez2018ethical, fiske2019your, knox2023justice, sedlakova2023conversational, akbulut2024all, wang2025understanding}. These issues include data privacy, the lack of regulatory frameworks, potential misuse that could exacerbate health inequalities, misinformation, toxic content, power dynamics, user autonomy, and risks from anthropomorphism, such as forming emotional connections with an AI tool.  These issues present serious ethical challenges for talk therapy, where a therapist's statements directly impact a client's therapeutic experience and where the therapist's agency to make decisions about client safety is key. Lastly, there has been increasing work highlighting how mental health apps fail to handle emergency suicidal crises and their inability to protect client data~\cite{tekin2021big,skorburg2022there,sedlakova2023conversational}. These concerns are particularly serious for populations historically excluded or marginalized by psychology, psychiatry, and psychotherapy, who face heightened risks from algorithmic biases embedded in AI systems~\cite{knox2023justice}.

In response, researchers are formulating guidelines for the responsible use of LLMs, emphasizing the need for an interdisciplinary approach to minimize potential harm and enhance transparency~\cite{stade2024large, xu2024mental}. Techniques like prompting~\cite{sun2025rethinking, sun2023human}, domain-specific fine-tuning~\cite{ji2022mentalbert}, and in-context learning are emerging as ways to align LLM outputs with psychotherapy principles while maintaining control over content~\cite{na2024cbt, sun2025rethinking, stade2024large}. One such example is the MentalBlend framework, which incorporates techniques from CBT and other psychotherapy frameworks to align LLM counselor behavior with professional standards~\cite{gu2023mentalblend}. 

\subsection{Human-AI Comparisons: Vignettes Are Not Enough}
Several psychotherapy-related studies have compared human responses to those of chatbots, indicating the promising effects of using the latter as a standalone solution. AI-generated responses are often perceived as more empathetic and helpful than those of professional experts in peer support~\cite{morris2018towards, young2024role}, talk therapy~\cite{vowels2024chatbots, hatch2025eliza}, and medical contexts~\cite{ayers2023comparing}. For instance, ~\citet{young2024role} evaluated blinded human- and AI-generated responses to help-seeking messages and found that participants preferred AI-generated responses over those of peer counselors. Similarly,~\citet{hatch2025eliza} compared GPT-4 responses to those of human therapists using vignettes on couples therapy, showing that participants could rarely tell the difference between the two. GPT-4's responses were also rated higher on psychotherapy principles. ~\citet{ayers2023comparing} analyzed 195 medical questions from Reddit's r/AskDocs and found that licensed healthcare professionals preferred LLM-generated responses over those of physicians in 78.6\% of cases, with chatbot responses rated significantly higher in both quality and empathy.

However, the above comparisons either:
\begin{enumerate}
    \item focus on individual, isolated responses (single-turn interactions) ~\cite{ayers2023comparing, morris2018towards, vowels2024chatbots} or 
    \item compare responses to \textit{therapy-like vignettes} that ``represent a hypothetical snapshot of therapy'' and may not reflect users' lived experience~\cite{hatch2025eliza}
\end{enumerate}
These methodological limitations make it difficult to generalize these positive findings regarding AI-generated responses to a model's behavior across an entire session.

To better understand how human and LLM counselors differ in their capacity to provide ongoing, context-aware support throughout a full session---one sensitive to the user's unique context and lived experience---a live experiment would seem necessary. However, researchers have raised concerns about experimenting with standalone LLMs on vulnerable populations. For instance, founders of a digital mental health company faced criticism for using LLMs in their peer support services without explicitly informing users, arguing that the nature of the test rendered it exempt from laws of informed consent. This approach was challenged by medical and technology experts, who questioned the harms the experiment could present~\cite{biron2023online}. Given the delicate nature of mental health support, deploying LLMs without a thorough understanding of the support they offer could be harmful~\cite{fiske2019ethical, xu2024mental, heinz2025randomized}. In their randomized controlled trial,~\citet{heinz2025randomized} manually monitored all chatbot responses to detect inappropriate content and contacted the participants to provide corrections. 

Hence, in this study, we made an intentional trade-off: instead of opting for a live experiment that may cause harm, we used publicly available, real-world session transcripts from peer counselors to generate matched sessions with LLM counselors. These synthetic sessions allowed us to avoid exposing vulnerable populations to potential harm. We then recruited clinical psychologists with expertise in CBT to evaluate CBT competence for each counselor.

\section{Methods}
\begin{figure}[ht]
    \centering \includegraphics[width=\textwidth]{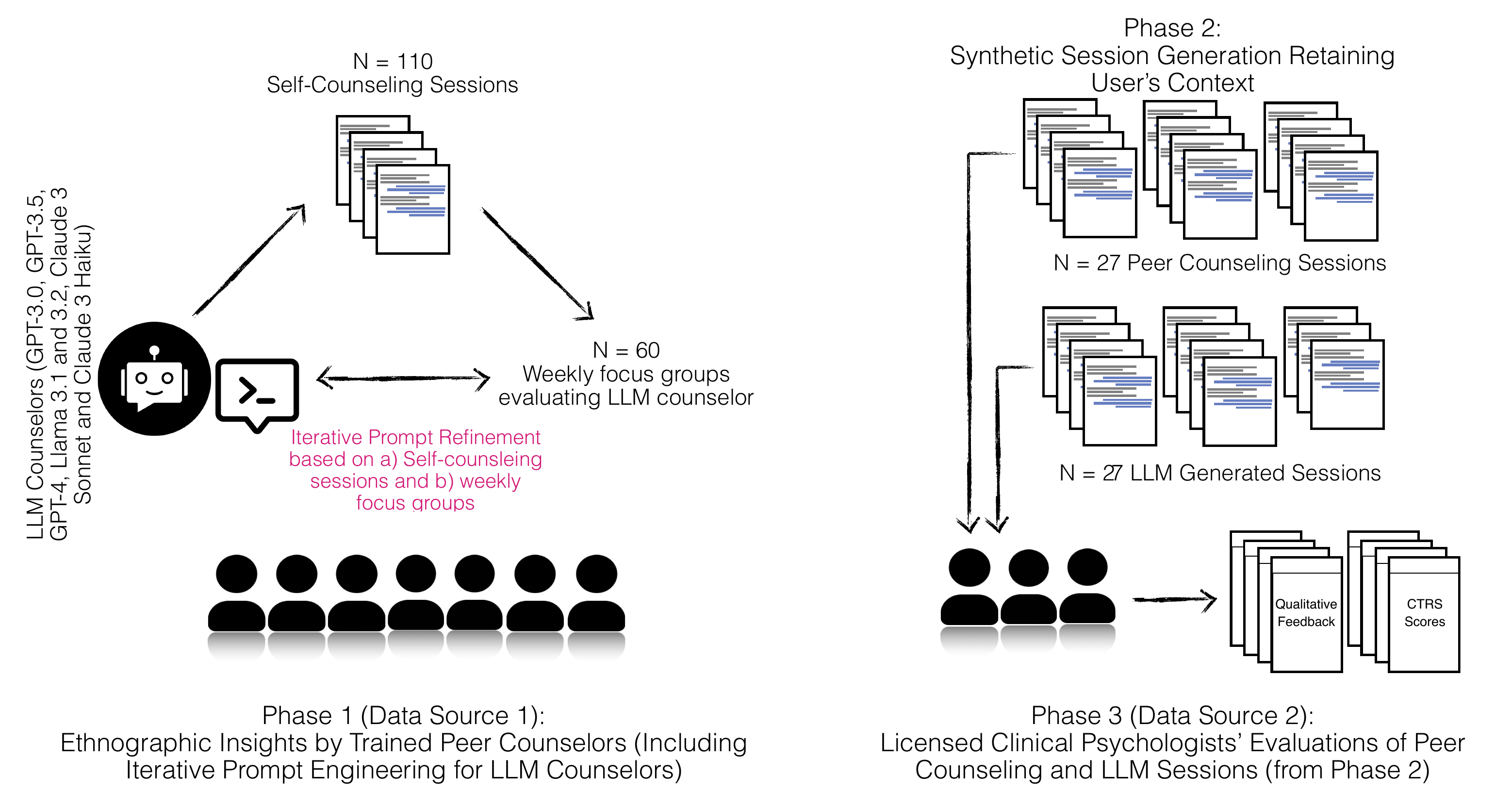}
    \caption{Data sources for the study. Over 18 months, CBT-trained peer counselors (N = 7) conducted 110 self-counseling sessions and 60 weekly focus groups while iteratively refining prompts across multiple LLMs (GPT-3.0, GPT-3.5, GPT-4, Llama 3.1 and 3.2, Claude 3 Sonnet, and Claude 3 Haiku). These ethnographic insights (Source 1) were triangulated with clinical evaluations (Source 2), in which three licensed clinical psychologists evaluated session-level CBT competence in both peer counseling and their LLM-generated counterparts using qualitative feedback and standardized CTRS scores.}
    \Description{Data sources for the study. Over 18 months, CBT-trained peer counselors (N = 7) conducted 110 self-counseling sessions and 60 weekly focus groups while iteratively refining prompts across multiple LLMs (GPT-3.0, GPT-3.5, GPT-4, Llama 3.1 and 3.2, Claude 3 Sonnet, and Claude 3 Haiku). These ethnographic insights (Source 1) were triangulated with clinical evaluations (Source 2), in which three licensed clinical psychologists evaluated session-level CBT competence in both peer counseling and their LLM-generated counterparts using qualitative feedback and standardized CTRS scores.}
    \label{fig:methods}
\end{figure}

The primary goal of our exploratory study was to understand and compare the behaviors of human and LLM counselors. To answer this, the core data source for our study comes from a real-time, online peer support platform where users engage in synchronous conversations with trained peer counselors. These human counselors form the basis of our comparison with LLM-based counselors; both were trained in CBT principles to provide single-session counseling. To address a key methodological challenge in AI therapy research, how to conduct comparisons between human and AI counselors without exposing vulnerable users to unvetted interventions~\cite{fiske2019ethical, biron2023online}, we used data triangulation (Figure~\ref{fig:methods}). The triangulation involved two data sources across three phases: (i) an 18-month remote ethnography with seven peer counselors (P4--P10), (ii) manual simulation of LLM counseling sessions using real multi-turn transcripts from a public mental health platform---segmented so that user utterances served as input to the LLM counselor to preserve users' lived experiences, and (iii) a comparative evaluation of a subset of sessions ($n_1 = 27$ human; $n_2 = 27$ LLM) by three licensed clinical psychologists (P1--P3).

\subsection{Phase 1: Ethnography with CBT-Trained Peer Counselors}
\label{ethnography}

\subsubsection{Study Context and Platform Overview:} For the first phase, we conducted a remote ethnography with seven peer counselors to understand how trained human counselors prompt LLMs with established CBT protocols and evaluate their clinical appropriateness. The counselors work on an online text-based peer support platform, called Cheeseburger Therapy, where users chat in real-time with a trained peer (who is called the human counselor) for single-session CBT-based counseling. The service has been active since 2017, as a research initiative to increase mental health accessibility by training laypeople in CBT techniques through a 25-40 hour training program.

The training process includes completing a custom CBT manual that teaches psychotherapeutic and empathetic techniques. including active listening, open-ended questions, reflective restatements, and cognitive reappraisal (identifying negative thought patterns to replace them with healthier beliefs). Upon completing the manual, peers participate in practice sessions with another peer in training. The site is maintained by a team including a Family Therapist, AI Research Scientist, UX Designers, Software Engineers, and peer counselors trained in CBT techniques. \textbf{The platform team has no overlap with the research team. The authors' role was purely observational and ethnographic.} 

\subsubsection{Prompting LLMs with CBT Principles:} 
Following the release of ChatGPT in 2022, peer counselors began experimenting with aligning various LLMs, including OpenAI's Generative Pretrained Transformer series (GPT-3, GPT-3.5, GPT-4), Llama 3.1 and 3.2, and Claude 3 Sonnet and Claude 3 Haiku, with CBT principles used in their training. Their goal was to explore whether LLMs could follow CBT techniques and lead counseling sessions, mirroring the approach taken by human counselors. Prompt engineering, as opposed to domain-specific fine-tuning, has emerged as a means to align LLM outputs with psychotherapy principles without extensive model retraining~\cite{rashkin2019towards, shah2022modeling, sun2023human, sun2025rethinking}.

Peer counselors developed a system prompt for GPT 3.0 (later updated to 3.5 and 4.0), instructing the LLM to follow the same CBT method outlined in their training manual used for human counselors. 
\newline
\newline

The prompt included key CBT principles, including:
\begin{enumerate}
    \item the Cognitive Model (explain to users how their thoughts influence their feelings and behaviors)
    \item Cognitive Reappraisal (guide users to identify and challenge their maladaptive thoughts)
\end{enumerate}

The CBT model assumes a reciprocal relationship between thoughts, feelings, and behaviors. A typical session flow is: thoughts \(\rightarrow\) feelings \(\rightarrow\) behaviors \(\rightarrow\) cognitive reappraisal.

\subsubsection{Systematic Testing Through Self-Counseling Sessions and Weekly Focus Groups:}
\label{self_counseling_sessions}
To evaluate the robustness of the prompt(s), peer counselors conducted $N = 110$ self-counseling sessions, acting as the client with the LLM. These self-counseling sessions were not persona-driven but were grounded in real scenarios from counselors' lived experiences. Counselors then met remotely on a weekly basis. All counselors were CBT-trained and used this training to discuss and evaluate the behavior of LLM counselors, identifying instances (both individual responses and overall session competence) where LLM counselors aligned or diverged from the CBT manual. We refer to these as the Peer Discussions Dataset, which forms our first qualitative dataset.

\subsubsection{Iterative Prompt Design Based on CBT Training:} Prompt design was iterative, informed directly by insights from these weekly focus group discussions. Peer counselors would iteratively edit the prompt to improve the model's ability to perform core CBT techniques. For instance, if peers identified that LLM counselors failed to ask for the user's values during the session, they would add an explicit instruction: \textit{``Ask the user about their values after they share their troubling thought.''} Each version of the prompt was tested in self-counseling sessions (Subsection \ref{self_counseling_sessions}), discussed in the weekly focus groups, and edited based on performance. 

After 18 months of evaluation (May 18, 2023--October 12, 2024), 110 self-counseling sessions, and 60 weekly focus groups, the counselors converged on a final version of the prompt for GPT 4.0 that they collectively deemed safe and consistent with CBT principles. While the team met weekly as a focus group, the first author attended 18 of these sessions (approximately one per month) and reviewed recordings of the remaining meetings. The final version of the prompt, recommended by peer counselors, was used to generate a subset (n = 27) of publicly available sessions. A summarized version of the prompt is provided in Appendix~\ref{append:prompt}

\subsubsection{Ethical Considerations (Phase 1):} The authors were not part of the peer counseling team and had no role in developing, editing, or evaluating the prompts for LLM counselors. Instead, our role was that of ethnographic observers. The first component of our two-source qualitative dataset was informed by these ethnographic observations gathered during remote meetings. We do not analyze the content of the self-counseling sessions; instead, this study focuses on the behavior of LLM counselors and how it aligns with or diverges from core CBT principles. Our study was determined to fall under program evaluation rather than human subjects research, as per our Institutional Review Board. However, in alignment with ethical guidelines and best practices, we took steps to safeguard participants, specifically by paraphrasing all quotes provided by counselors to ensure confidentiality.

\subsection{Phase 2: Generating Sessions While Retaining a User's Lived Experience}\label{session_simulation}

The methodology of this phase is motivated by the shortcomings of two common practices in simulating and evaluating client-LLM sessions: (1) authors posing as potential clients~\cite{aleem2024towards, hatch2025eliza, vowels2024chatbots} or (2) LLMs acting as clients given a user's publicly available session transcript~\cite{chiu2024computational}. First, when authors pose as clients, they can reflect their own biases (e.g., interviewer bias, confirmation bias) in simulating clients. This approach also assumes that, given a certain user persona, the simulated client would have predetermined key negative thoughts and behaviors. For instance, in their simulated sessions,~\citet{aleem2024towards} assumes a person from Pakistan living in the USA would struggle to communicate with their family. Similarly,~\citet{hatch2025eliza} created counseling vignettes based on the author's personal clinical experience. \textbf{While these simulations can highlight frequently observed patterns of negative thoughts and behaviors across certain populations, they do not accurately capture the true context of a user. The thoughts assumed in these vignettes are merely broad trends, not personal truths.} Second, when relying on LLMs to simulate clients~\cite{chiu2024computational}, the sessions can suffer from limitations observed in LLMs' behaviors, such as hallucination, which can introduce the model's biases when simulating users' lived experiences~\cite{xu2024mental}.

\subsubsection{Generating Sessions Using User Messages as Input Prompts:} To address these issues, we generated LLM sessions by holding the user side constant, retaining the user's lived experience and allowing for a direct controlled comparison. We used the CBT system prompt, the final version for GPT 4.0 agreed upon by the peer counselors at the end of \hyperref[ethnography]{Phase 1}. Since CBT is an ``active, directive, time-limited and structured approach''~\cite{beck1976cognitive}, the sessions are likewise structured and typically follow a consistent protocol: (1) counselors identify a triggering event, (2) ask for the associated negative thoughts, emotions, and behaviors, and (3) apply cognitive reappraisal to challenge and reframe the client's negative thinking patterns. This structured, timely format has made CBT suitable for many internet-based interventions~\cite{morris2015efficacy, rennick2016health, fitzpatrick2017delivering, smith2021effective, prochaska2021therapeutic}. 

\begin{figure}
    \centering \includegraphics[width=\textwidth]{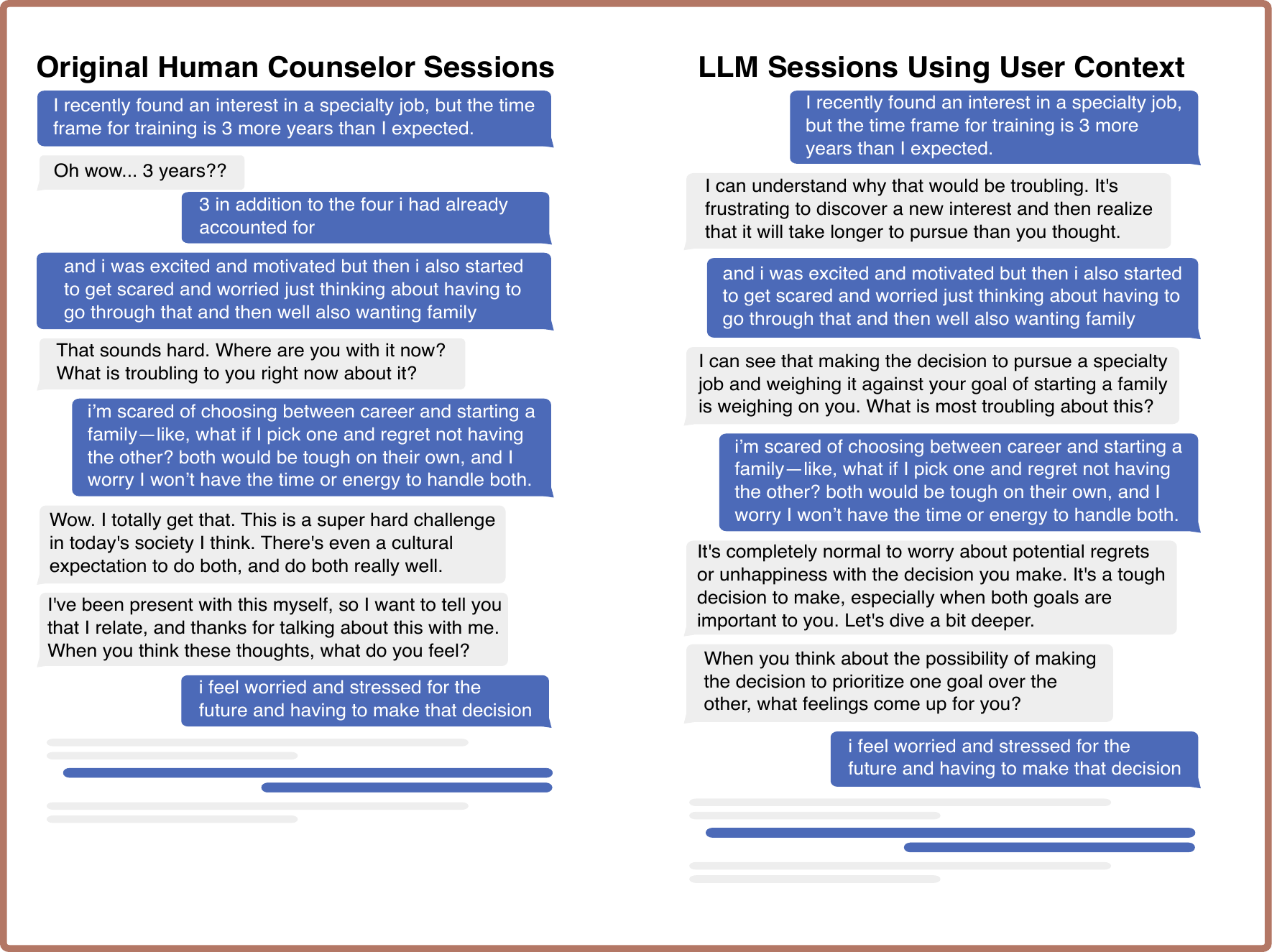}
    \caption{Synthetic Session Generations with a CBT-Prompted LLM While Retaining a Client's Lived Experience: Client's messages from original counseling sessions were used as input prompts to a CBT-prompted LLM counselor that follows a structured CBT protocol (thoughts \(\rightarrow\) feelings \(\rightarrow\) behaviors \(\rightarrow\) cognitive reappraisal), enabling a direct controlled comparison with human counselors.}
    \Description{Synthetic Session Generations with a CBT-Prompted LLM While Retaining a Client's Lived Experience: Client's messages from original counseling sessions were used as input prompts to a CBT-prompted LLM counselor that follows a structured CBT protocol (thoughts \(\rightarrow\) feelings \(\rightarrow\) behaviors \(\rightarrow\) cognitive reappraisal), enabling a direct controlled comparison with human counselors.}
    \label{fig:session_simulation}
\end{figure}

Clients' messages from the original human counseling sessions were used as input prompts to the CBT-prompted LLM to generate a subset (n = 27) of publicly available CBT sessions. First, two authors individually read the original human counseling sessions to understand and extract the key client responses that captured their context, focusing on their emotional and cognitive details (triggering events, their thought patterns, their feelings and behaviors, and any other specific life struggles the client explicitly mentioned during the session). After segmenting the dialogue into multi-turn utterances, we synthetically generated the sessions with the LLM counselor using the transcript of the original sessions, where the utterances spoken by the client were provided as the input prompt to the LLM counselor. The LLM counselor would then generate a response (following the CBT protocol) as a pseudo-counselor as shown in Figure~\ref{fig:session_simulation}.

For instance, during the session, if the LLM counselor asked: \textit{When you think ``My new job will disrupt my plan for family and other goals,'' how does that make you feel?} (thought \(\rightarrow\) feeling), we used the user's original message as the input prompt: \textit{``i feel worried and stressed for the future and having to make that decision''}. By providing each LLM output response with the next corresponding input response(s) (user message(s) from the original session), we preserved (a) the natural CBT sequence that both human peers and LLM counselors were trained to follow and (b) the context and lived experience of the actual user from the original session, which is often overlooked when simulating interactions. However, if an LLM did not probe deeper into a user's response, we excluded the subsequent user message from the session alignment. For instance, in the human counseling session, the counselor responded to the client's disclosure—``the time frame for training is 3 more years than I expected''—with a follow-up (``Oh wow... 3 years??'') that elicited further elaboration from the client (``Three in addition to the four I had already accounted for''). Since the LLM counselor did not produce a comparable prompt or probe, the corresponding elaboration was not used.

\subsubsection{More Than A Vignette: Session Recreations Grounded in User Context:} By holding the user side constant, we achieved a controlled comparison of human and LLM counselor behaviors without exposing vulnerable users to unvetted interventions. Both human and LLM counselors responded to identical user inputs while following the same CBT techniques and the structure of a single-session intervention (thoughts \(\rightarrow\) feelings \(\rightarrow\) behaviors \(\rightarrow\) cognitive reappraisal). Generating LLM sessions this way achieved two outcomes: (a) retained users' lived experience especially their key negative thought, feelings, and behaviors--elements that demand to be effectively addressed for successful CBT intervention (Figure~\ref{fig:session_simulation}), and (b) allowed for a controlled comparison of human- and LLM- counselors' behaviors. The only difference between the two session types was the counselor: human or LLM.

Two authors then reviewed each other's LLM-generated sessions. We had detailed conversations within the research team to ensure the sessions reflected the original. However, because we were familiar with the research question, this approach could introduce confirmation bias and interviewer bias. To mitigate such unintentional bias, we shared a subset of the sessions with peer counselors (from \hyperref[ethnography]{Phase 1}) to confirm the LLM-generated sessions reflected users' lived experiences from the original sessions.

\subsubsection{Ethical Considerations (Phase 2):} Unlike live studies requiring new participant engagement (and the risk that entails), our method uses publicly shared, consented counseling data to evaluate sessions with minimal risk to individuals while still enabling comparative analysis of human and LLM counselor behavior. As the peer support platform is entirely online and based on text-only communication, some of the historical human counseling sessions are publicly available online. These sessions were made public after users provided written consent for public sharing, obtained only after the session had ended. Users were informed that sharing was voluntary, and only sessions with explicit consent were included in this study. None of the sessions contained personally identifiable information (PII).

\subsection{Phase 3: Psychologists' Evaluations to Compare Human and LLM Counselors}

\label{session_evaluation}
To evaluate and compare human and LLM counselor behaviors from a clinical perspective, we shared both session types ($n_1 = 27$; $n_2 = 27$) with three licensed clinical psychologists. The goal of these expert evaluations was threefold: (a) to understand the behavior of LLMs from a clinician's perspective, (b) 
to compare human and LLM counselors with input from domain experts not involved in prompt development in \hyperref[ethnography]{Phase 1}, and (c) to triangulate data sources. The psychologists were not informed which sessions were conducted by humans or LLMs. They rated sessions independently to prevent any potential bias from influencing one another's assessments.  The following measures were used to rate each session:

\subsubsection{Cognitive Therapy Rating Scale (CTRS):} The Cognitive Therapy Rating Scale (CTRS), also known as the Cognitive Therapy Scale (CTS), is an 11-item instrument measuring counselor competence in administering CBT~\cite{young1980cognitive, vallis1986cognitive}. CTRS is the most common and widely used measure of CBT competence~\cite{muse2013systematic}, and has been used as a benchmark for evaluating CBT fidelity in clinical trials~\cite{borkovec2002component} and training programs~\cite{keen2008assessing, creed2016implementation}. CTRS is organized into two subscales: \textit{General Therapeutic Skills} and \textit{Conceptualization, Strategy, and Technique}. The former evaluates a counselor's ability to establish a strong therapeutic relationship, while the latter assesses CBT-specific skills. Each subscale consists of individual items to quantify a particular skill. 
Each item on CTRS is rated on a 7-point Likert scale ranging from 0 (incompetence) to 6 (excellent), with higher scores indicating greater competence. Total scores range from 0 to 66, with scores of 40 or above indicating good CBT competence~\cite{young1980cognitive, shaw1999therapist}.

\subsubsection{Session Feedback Surveys:} To complement the quantitative CTRS ratings with qualitative insight, we asked psychologists to complete two surveys: the \textit{Session Feedback Survey} after each individual session and the \textit{Counselor Comparison Survey} once they had reviewed both the original human session and its LLM-generated counterpart. 

We first asked psychologists to reflect on each session by focusing on counselor behaviors (See Appendix~\ref{append:session_feedback}), what worked well, and where improvements could have been made. The session-specific survey helped us elicit their reflections regarding each particular session. Next, we asked psychologists to compare the behaviors of both counselors by completing the \textit{Counselor Comparison Survey} (See Appendix~\ref{append:session_comparsion}) to (a) identify any unique insights or actions from each counselor that the other might not have provided and (b) assess which counselor demonstrated a better application of CBT techniques. This approach was intended to highlight the unique strengths and limitations of human versus LLM counseling. Lastly, we conducted a semi-structured interview with all three psychologists, lasting thirty minutes, focused on their reflections following the session evaluations.

\subsection{Design Considerations \& Trade-offs}

In designing the study methodology, we made a deliberate trade-off to avoid patient harm, which affects how our findings should be interpreted. Rather than simulating interactive dialogue, our experimental design uses a controlled generation of CBT-structured sessions, in which the client side is held constant. This setup allows a direct comparison of how different counselors (human or LLM) respond to an identical `user'. While current approaches to session simulations afford higher interactional fidelity, they provide limited experimental control for isolating session-level characteristics and are subject to author and confirmation bias. In contrast, fixing client inputs reduces the interactional nature but affords greater control to compare counselors' behavior, therapeutic framing, and CBT adherence under matched conditions.

However, this design involves additional trade-offs. Our setup for generating sessions with the CBT-prompted LLM counselor was highly constrained on the original transcripts. While this helped us preserve the lived experience of the user from original sessions, it is \textit{likely} that LLM-generated sessions, if re-conducted with the same user, would have diverged from the original transcripts, as users might have responded differently to the LLM counselor's responses. This divergence is not captured in our design. Recognizing the constraints of the data produced in (\hyperref[session_simulation]{Phase 2}), we do not quantitatively analyze or report CTRS scores. 

Lastly, to frame our ability to make qualitative claims about features like \textit{Collaboration}, \textit{Pacing}, and \textit{Guided Discovery}, we clarify our engagement with them as specific therapeutic features of the counselor's performance, and not as purely emergent conversational dynamics. While collaboration and pacing broadly depend on the engagement between individuals, it is important to note that this study engages with them from the perspective of the counselor. The observations made by the psychologists in \hyperref[session_evaluation]{Phase 3} were not broadly about the presence or nature of collaboration within the session, but rather the counselor's ability to have created that collaboration or to have paced the session. For example, \textit{Collaboration} and \textit{Pacing} can be functionalized as questions like \textit{``did the counselor ask what the user thinks would be helpful?''} and \textit{``did the counselor spend time unpacking what the user disclosed before jumping to cognitive reappraisal?'}' both of which are not centrally about emergent dynamics.

\subsection{Data Analysis}

To analyze our qualitative data, we used an integrated analytic approach~\cite{o2021mixing}, combining ethnographic observations from \hyperref[ethnography]{Phase 1} with clinical evaluations from \hyperref[session_evaluation]{Phase 3}, treating them as complementary rather than separate data sources.

We used a combination of inductive and deductive inquiry to iteratively refine our coding framework~\cite{o2021mixing}. We chose this approach because our qualitative methods were designed to complement each other, incorporating perspectives from both peer counselors and licensed clinical psychologists. This enabled a deeper engagement with our research objectives, helping us identify patterns of inconsistency and similarity across contexts through critical comparison, consistent with grounded theory~\cite{glaser1999discovery}. 


We used the constant comparative method to code and generate themes from our data~\cite{glaser1965constant}. 
In the following, we outline the key assumptions that shaped our analysis~\cite{o2021mixing}. Given that our study explored an under-researched area of human vs. LLM counseling, our analysis was neither strictly data-driven (inductive) nor completely theory-driven (deductive). On one hand, we were driven by theory, as our understanding of counseling practices and the items listed in the CTRS guided our initial coding framework. However, as we progressed, we identified additional themes emerging from the data that were not encompassed by the existing CTRS items, and these challenged the emerging theory, inviting a process of reflection and comparison with other parts of the data. In line with~\citet{glaser1965constant}, we sought to maintain an integrated, consistent, and plausible theory grounded in the data, without letting rigid coding hinder new insights, so that theory was constantly being created and refined in a systematic way.

Through regular collaborative discussion, we organized emerging codes into larger themes, an iterative process that evolved as we analyzed more data. These themes included, for example, consistent patterns in how peer counselors conveyed empathy compared to LLMs, or ways in which both counselors struggled differently with leading CBT sessions. Preliminary themes were also reviewed with the second author, who is a clinical psychologist. Our data analysis allowed us to construct a grounded yet theory-informed understanding of effective counseling behavior (RQ1), counseling challenges (RQ2), and the implications for human–AI collaboration in designing counseling interventions (RQ3).

\subsection{Data Release} To enable replication and further research, we publicly release\footnote{https://github.com/brownhci/human-llm-cbt-evals} the original human counseling sessions and LLM sessions from \hyperref[session_simulation]{Phase 2}, and the Psychologists Evaluation Dataset from \hyperref[session_evaluation]{Phase 3}. Each session is associated with message and session-level attributes, including a ``Session ID'' (text), ``Counselor ID'', and the ``Source'' of the session (binary), indicating whether the session was conducted by a human counselor or generated with the LLM counselor. The Message Attributes include the content of the ``Message'' (text), a binary field ``FromThinker'' to determine if the message is sent by the user or counselor, ``Timestamp'' (e.g., Sat Feb 19 2022 17:49:52 GMT-0500), ``Offset'' (e.g., Eastern Standard Time), and ``MessageID''. The Session Notes Attributes record CBT-based labels for each session, such as ``Event'', ``Thoughts'', ``Feelings'', ``Behavior'', and ``Cognitive Distortion''. Sessions are accompanied with the ``Session Feedback'' (text) provided by all three psychologists. For a detailed dataset schema, refer to Appendix~\ref{append:session_schema_1} and Appendix~\ref{append:session_schema_2}.


\section{Findings}
In this study, we compared the session-level behaviors of human peers and LLM counselors both trained to provide single-session CBT-based counseling. \cscwfinal{In Subsection~\ref{quant_findings}, we examine CTRS item-level behaviors descriptively, drawing on ratings from the three licensed psychologists (P1--P3).} In Subsections 4.2 and beyond, we triangulate our results by looking deeper into the qualitative data collected during the ethnographic observations (\hyperref[ethnography]{Phase 1}), session examples (\hyperref[session_simulation]{Phase 2}), and session evaluations (\hyperref[session_evaluation]{Phase 3}). Throughout, CTRS scale items are referenced \textit{in italics} within parentheses (e.g., \textit{Interpersonal Effectiveness}).

\subsection{Different Paths in CBT Practice: Relational Conversation vs Algorithmic Structure}
\label{quant_findings}

\newcommand{\ra}[1]{\renewcommand{\arraystretch}{#1}}
\begin{table*}
\centering
\ra{1.55}
\setlength{\tabcolsep}{9pt}
    \caption{Comparison of Human Peers and LLM Counselors' Session-Level Behavior as Measured by the CTRS Scale, Across General Therapeutic Skills, CBT-Specific Techniques, and Risk for Patient Harm}
        \begin{tabular}{@{}>{\raggedright\arraybackslash}p{3cm} >{\raggedright\arraybackslash}p{5cm} >{\raggedright\arraybackslash}p{5cm}@{}}
        \toprule
        \textbf{Behavior} & \textbf{Human Counselor} & \textbf{LLM Counselor} \\
        \midrule
        \multicolumn{3}{c}{\textbf{General Therapeutic Skills}} \\ \midrule  
        \textbf{Agenda}
        & Neglects agenda-setting at times 
        & Sets a clear agenda\\
        \textbf{Feedback}
        & Asks well-crafted open-ended questions throughout the session
        & Must be manually~interrupted by the user for feedback\\ 
        \textbf{Understanding}
        & Interprets implicit cues; reflects users' lived experience through nuanced verbal acknowledgment
        & Tends to treat all client disclosures as equally important, lacking the clinical judgment to triage what matters\\
        \textbf{Interpersonal} 
        & Uses relationship building strategies (small talk and self-disclosure) to build trust and therapeutic alliance
        &
        Cold and formulaic; however, imparting subjective interpersonal skills would be deceptive \\
        \textbf{Collaboration}
        & Active; encourages the user to take an active role during the session & Passive; tells the user what to do instead of co-creating solutions\\ 
        \textbf{Pacing} 
        & Off-topic digression (engages in non-CBT discussion) & Stays on topic but frequently rushes applying the method \\ \midrule
        \multicolumn{3}{c}{\textbf{Conceptualization, Strategy, and Technique}} \\ \midrule
        \textbf{Guided Discovery}
        & Uses skillful questioning to guide self-reflection and help the user draw their own conclusions & Lectures the user into change through imposed solutions; advice-giving rather than guided discovery\\ 
        \textbf{Focusing on Key Cognition}
        & Good demonstration but can focus on too many thoughts at once 
        & Good~demonstration~but~can latch onto~surface-level maladaptive thoughts\\ 
        \textbf{Strategy for Change}
        & Adapts CBT~strategies~to~client's values and context,~but~inconsistently 
        & Follows a protocol-driven strategy but applies it inflexibly regardless of client context\\
        \textbf{Application of CBT} 
        & Relational; applies CBT techniques effectively, but sometimes engages in unnecessary self-disclosure that distracts from session's focus
        & Technical; applies psychoeducation effectively but is narrow in its approach and requires human-in-the-loop for relational intervention\\ 
        \midrule
        \textbf{Harms to Patients} & 
        Fails to detect key mental~health cues (e.g., depression); risks engaging in non-evidence-based techniques (e.g., inner child work, venting without structure) & 
        
        Over-validates unhealthy beliefs; abruptly abandons users disclosing sensitive topics; deceptive empathy through anthropomorphic design\\ 
        \bottomrule
    \end{tabular}
\end{table*}

Despite using the same CBT techniques, human peer and LLM counselors took markedly different therapeutic approaches to directing and guiding sessions.

Human counselors used relational techniques that centered the therapeutic relationship and collaborative exploration as central tools for change. Counselors used self-disclosure and small talk (\textit{Interpersonal Effectiveness}) to build rapport and trust. They were more skilled at recognizing implicit cues (\textit{Understanding}) and asking open-ended questions (\textit{Feedback}) beyond the cognitive model to understand users' lived experiences within their broader cultural, religious, and personal contexts. All participants (seven peer counselors and three licensed clinical psychologists) noted that these relationship-building strategies led users to take an active role during the session, allowing greater room for exploration, but at the risk of losing therapeutic focus.

However, extensive use of relationship-building strategies (including non-therapeutic discourse such as self-disclosure and small talk) on topics beyond CBT had mixed reviews: they were sometimes seen as strategies for building therapeutic alliance, but were at times deemed unnecessary and unstructured (\textit{Agenda}), without adding much value to the session, reflecting the low psychologist agreement (\textit{Interpersonal Effectiveness}). Counselors also sometimes struggled with therapeutic focus in other parts of the session, for example, they would attempt to address too many thoughts or behaviors at once (\textit{Focusing on Key Cognition \& Behaviors}).

On the other hand, LLM sessions were more aligned with technical CBT intervention. These sessions were more structured (\textit{Agenda}) and had higher levels of psychoeducation, incorporating specific cognitive-behavioral techniques such as suggesting breathing and mindfulness techniques (in cases where a user was having a panic attack) (\textit{Strategy for Change}). The LLM counselor quickly focused on specific cognitions or behaviors relevant to the user's problem (\textit{Focusing on Key Cognition \& Behaviors}) but had trouble identifying the more central cognitions or behaviors that \textit{``offered greater promise for progress''} (P1) (\textit{Understanding}). This also manifested as a difficulty separating clinically important disclosures from trivial ones. The LLM counselor's ability to pace a session received mixed reviews. While LLM counselors stayed on track, they were rigid in their approach and rushed into applying the method, at times tending to lecture the user and impose solutions rather than collaboratively guide them.

Overall, we found a dynamic tension between relational engagement and session structure in the counselors' \textit{Application of CBT}. Given that human peer counselors were stronger in relational engagement and collaborative exploration, but inconsistent in therapeutic structure and at times unfocused in addressing key cognitions, our findings point to human peer counselors as delivering a moderately effective but structurally uneven intervention. In contrast, since LLM counselors were rigid in approach, less able to distinguish clinically important content from trivial disclosures, and more likely to lecture and impose solutions rather than collaboratively guide users, our findings point to LLM counselors as delivering a less effective intervention overall.

\subsection{On Relational–Structural Dynamics in Counseling}

\begin{figure*}[b!]
    \centering
    \begin{subfigure}[t]{0.5\textwidth}
        \centering
        \includegraphics[height=1.3in]{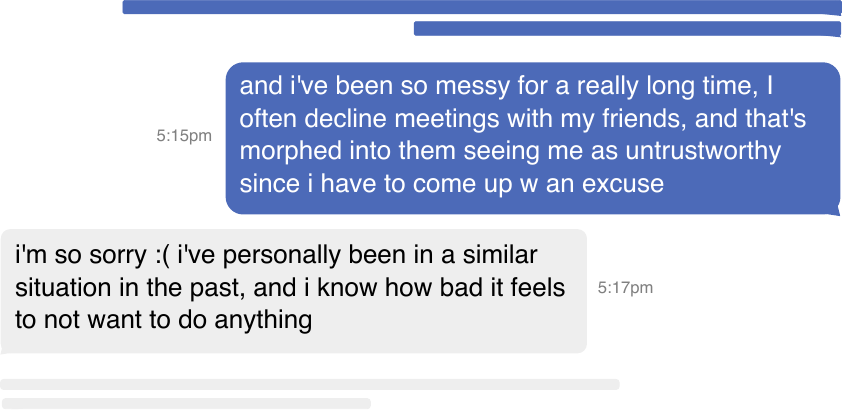}
        \caption{Original Human Counseling Session}
        \Description{Original Human Counseling Session}
    \end{subfigure}%
    ~ 
    \begin{subfigure}[t]{0.5\textwidth}
        \centering
        \includegraphics[height=1.45in]{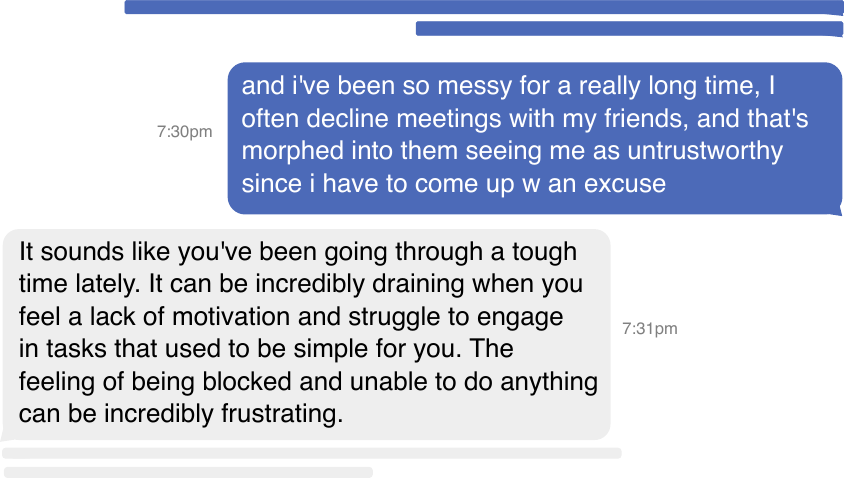}
        \caption{LLM-Generated Counseling Session}
        \Description{LLM-Generated Counseling Session}
    \end{subfigure}
    \caption{Human counselors rely on relational strategies, while LLM-based counselors use CBT techniques to validate clients}
    \Description{Human counselors rely on relational strategies, while LLM-based counselors use CBT techniques to validate clients}
\end{figure*}

\subsubsection{Human Counselors Focus on Small Talk and Self-Disclosure}

Counselors used a variety of relational conversational strategies, including small talk and self-disclosure, to establish a therapeutic alliance. The sessions were characterized by \textit{``authentic rapport''}, often including \textit{``random chatter, which led to deeper self-reflection and self-disclosure''} (P1). Counselors (P4--P10) noted how self-disclosure helps clients feel heard and understood, a behavior that was also observed by psychologists (P1--P3) during their independent session evaluations. For example, in the original counseling session where a client described feeling isolated due to unproductivity, psychologists noted that the peer counselor's self-disclosure was more impactful than standalone CBT techniques.

    \studyquote{The peer counselor was able to generate a tremendous amount of credibility by self-disclosure, connecting their own experiences with those of the client.}{1}

In addition, peer counselors also used empathetic paralinguistic expressions and rapport-building statements (similar to nonverbal cues like nodding or ``mhm'' in face-to-face conversation) such as ``ooof'' or ``Oh wow...3 years'' (Figure~\ref{fig:session_simulation}) as listening cues. P3 noted: \textit{When you use statements like ``Oh, that's nice'' or ``Oh wow...3 years?'', it can make the client feel heard.}


\cscwmajorrevision{While in some cases, this  \textit{``additional therapeutic chatter added to session quality''} (P1) by establishing rapport or gaining context, at other times, psychologists noted that it detracted from the session's therapeutic goals, as human counselors often (a) failed to begin sessions with a clear, structured agenda, (b) went off-track and engaged in non-CBT discussions (off topic remarks, such as taking a break to feed a neighbor's dog during a session) or (c) focused on too many parts of the sessions at once, all of which affected the pacing of the session (\textit{Pacing}) and \textit{``did not match CBT fidelity''} (P2). P3 noted how \textit{``exploring additional aspects unrelated to the session, without the client's consent, could hurt rapport''}.}



\subsubsection{LLM Counselors `Paint by Numbers' with a Focus on Psychoeducation}

In contrast to human counselors, LLM counselors communicated exclusively through CBT techniques, resulting in structured but relationally limited sessions. \cscwmajorrevision{Positive feedback for these sessions centered on agenda-setting, pacing, and adherence to CBT conceptualization, with comments noting that the LLM counselor \textit{``kept the session organized, structured and CBT-focused''} (P2), \textit{``did well with offering validation and active listening responses while also re-framing negative thoughts''} (P9), and \textit{``brought in cognitive behavioral techniques such as taking a brief time-out from interaction and using breathing techniques to help the client manage their emotions''} (P3).}

However, with this ascribed focus on CBT concepts, LLM counselors, by extension, were more passive agents within sessions, only able to interpret what was explicitly prompted either in the CBT system prompt (\hyperref[ethnography]{Phase 1}) or through explicit input prompts provided from the original human counseling sessions (\hyperref[session_simulation]{Phase 2}). P3 described this close adherence to standard CBT protocol as a starting foundation of counselor training, describing: 

    \studyquote{At first, a counselor needs to learn to \textbf{paint by numbers}, e.g., stick to the treatment manuals and learn the exact way CBT techniques should be implemented. Later, humans (especially more advanced counselors/therapists) incorporate their personality and become more flexible and creative, while still maintaining the core of the CBT techniques.}{3}

\cscwmajorrevision{P3 furthered that the end goal was to \textit{``do it automatically,''} insofar as to then \textit{``paint freehand,''} adding in more personality. The inability to use relational techniques without being deceptive was a major barrier for LLM counselors. As P1 explained, \textit{``a lot of times, human counselors are going to have self-referential statements, even if not necessarily self-revealing. The problem is that for AI to be self-referential, it would necessarily be deceptive.''} This tension was also reflected in ethnographic observations, where peer counselors described LLMs' attempts at empathy as \textit{``humanizing an experience that is not human.''} (P5)}

It is worth noting that while prompting LLM counselors to follow CBT techniques, the peer counselors also went back and forth in giving the chatbot a persona, but ultimately decided not to. This design decision was intentional: to understand what human counselors bring to a support session, that (a) cannot be replicated and (b) would be unethical to automate.



\subsection{On Understanding User's Lived Experience}

\subsubsection{``I Get That'': Human Counselors Use Value-Centered CBT Approach}

\cscwmajorrevision{Human counselors asked intuitive, open-ended questions throughout the session to gather context that helped them (a) build a holistic understanding of the client's lived experience and (b) tailor CBT techniques to the client's situation. Psychologists (P1-P3) emphasized that the ability to go beyond the cognitive model (which focuses narrowly on thoughts, feelings, and behaviors) enabled counselors to tailor interventions by incorporating broader aspects of the client's context, including cultural and religious factors. As P5 explained, \textit{``part of the challenge is going beyond the words expressed by the client''.}} 

This extended to noticing implicit cues that went beyond what clients explicitly stated. For instance, during one session where a client was struggling with panic attacks, the counselor explored whether these were frequent or recent. Through additional context, the counselor interpreted that the client had recently become a father and was experiencing additional cultural expectations associated with that role. This insight allowed the counselor to adapt CBT \textit{``by identifying appropriate thoughts that respected the cultural and religious milieu the client came from''} (P1). 


\cscwmajorrevision{A particular strength was counselors' focus on clients' values to guide the session. Psychologists noted that this focus on values was particularly impactful, emphasizing that the \textit{``most effective counselors discussed the client's values and then tied these values into identifying thinking errors''} (P1). For example, in one session, a client from the Global South faced familial and cultural pressures not typically present in Western societies. Rather than generalizing these experiences, human counselors empathized and acknowledged the impact of upbringing on difficult conversations, stating, ``culture and upbringing can have a way of making it difficult to talk about things''. They also drew from their own parallel experiences, expressing understanding: ``my husband also comes from a culture where he is expected to provide.''}

\subsubsection{``I Almost Get It Too'': LLM Counselors Struggle with Values and Implicit Cues}

\cscwmajorrevision{LLM counselors demonstrated a shallow understanding of users' lived experiences. This issue stemmed from several limitations: they were unable to ask intuitive questions beyond the CBT framework, lacked cultural sensitivity, tended to over-generalize users' experiences, and relied on a one-size-fits-all adaptation of CBT techniques. As a result, LLM responses \textit{``felt disconnected from the client's internal reality''} (P1), often defaulting to excessive, text-heavy summaries that read ``as if the task were to summarize literally everything the client said'' (P9). As one peer counselor noted, the model \textit{``is too focused on the wrong thing''} (P6), while another observed that it ``misses being intelligent on toning down what really matters to a client'' (P7).}

This also manifested as an inability to distinguish between what was most important in a client's narrative and what was less central. For example, in one session, a user mentioned several connected details over time: experiencing panic attacks, having a child, and cultural values. The human counselor drew on intuition to go beyond the training manual, asking whether the child was a recent addition, what cultural expectations were at play, and how these factors related to the user’s panic attacks. In contrast, the LLM counselor did not probe for this additional context, failing to explore how these details might be meaningfully connected.

\cscwmajorrevision{This difficulty in prioritizing and connecting client information was closely tied to the LLM's inability to move beyond a rigid, procedural CBT structure, which emerged as one of the most noticeable limitations during sessions. Reflecting on a self-counseling interaction, one peer noted:}

    \studyquote{As a counselor, instead of immediately diving into CBT, you start by asking the client what's currently going on in their life and what they're trying to move towards. What are the things that truly matter to them? What areas of their life are we addressing:  career changes, relationship challenges? The [self-counseling] session with the LLM was shallow because it didn't explore what truly matters.}{6}

\subsection{On User Collaboration and Guided Discovery}

\begin{figure*}[b!]
    \centering
    \begin{subfigure}[t]{0.5\textwidth}
        \centering
        \includegraphics[height=2in]{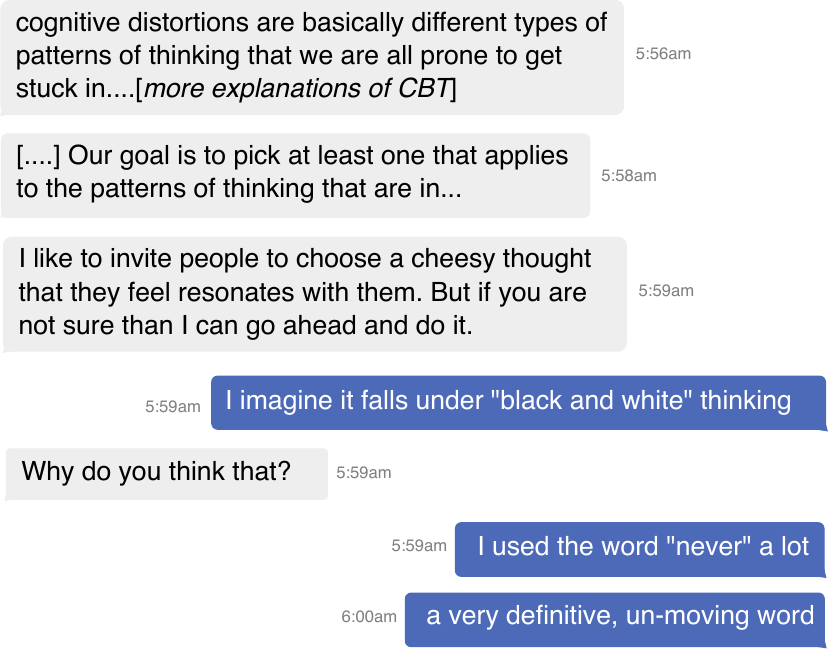}
        \caption{Original Human Counseling Session}
        \Description{Original Human Counseling Session}
    \end{subfigure}%
    ~ 
    \begin{subfigure}[t]{0.5\textwidth}
        \centering
        \includegraphics[height=2in]{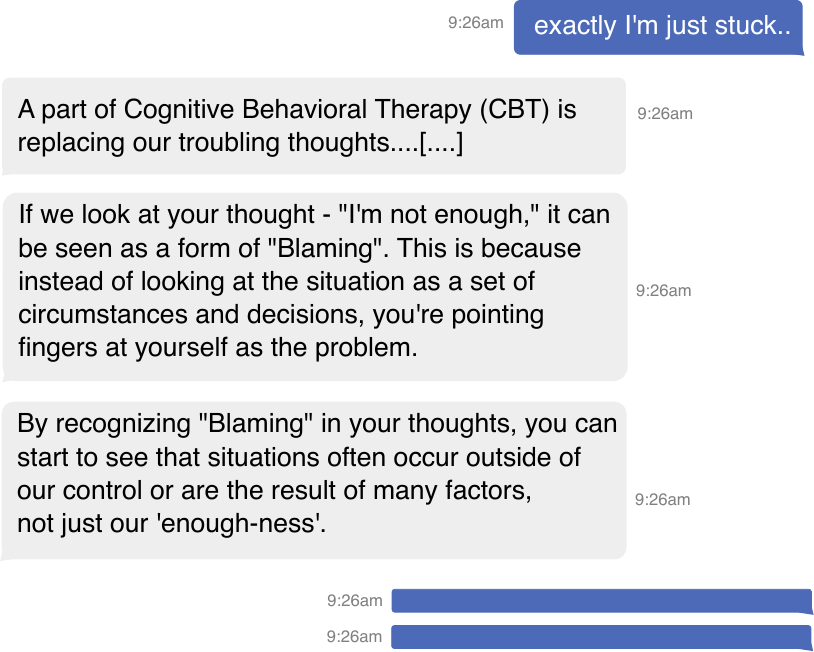}
        \caption{LLM-Generated Counseling Session}
        \Description{LLM-Generated Counseling Session}
    \end{subfigure}
    \caption{Human counselors using reflective questioning to help the client identify their own cognitive distortions whereas the LLM counselor tended to zero-in on thinking errors without seeking user collaboration}
    \Description{Human counselors using reflective questioning to help the client identify their own cognitive distortions whereas the LLM counselor tended to zero-in on thinking errors without seeking user collaboration}
    \label{collaboration}
\end{figure*}

\subsubsection{Human Counselors' Encouraged Active Participation:}

Human counselors encouraged clients to take an active role during the session. They nudged clients to identify their cognitive distortions, discuss why certain ones were relevant and others were not, and generate alternative healthy thoughts, offering support when clients struggled to come up with them independently. P3 annotated the following excerpt from an original peer counseling session (Figure~\ref{collaboration}) where the counselor's collaborative skills were particularly strong:

\studyquote{The counselor used excellent probing and reflective techniques to encourage the client to reexamine his own thoughts.}{3}

This collaborative approach extended to pacing: human counselors checked in with clients before moving forward, invited them to choose and apply cognitive distortions to their situations, gave them space to process, and asked for client feedback before transitioning to cognitive reappraisal. Rather than presenting solutions, counselors invited clients to generate their own alternative thoughts, stepping in with guidance only when needed. This approach reflected a \textit{``shared, iterative discovery process rather than a one-directional intervention''} (P1).

\subsubsection{LLM Counselors' Psychoeducation Becomes Lecturing and Gaslighting:}

LLM counselors treated users as passive participants, telling clients what to do and how to fix their problem. Counselors would spoon-feed CBT psychoeducation in bulk through verbose responses, with little to no give-and-take from the client. There were instances of text heavy responses that would intellectualize clients' messages, \textit{``as if the task were to summarize literally everything''} (P1). \cscwmajorrevision{Similarly, one peer observed that the LLM \textit{``speeds through the method without any back-and-forth, with its text-heavy responses''} (P6).}

The lack of collaboration and guided discovery became more pronounced toward the end of sessions. LLM counselors over-relied on a single cognitive distortion by \textit{``zeroing in on a thinking error''} (P2), without nudging the client to identify their own distortions or consider alternative thoughts. \cscwmajorrevision{P4 noted that toward the end, the LLM \textit{``immediately addresses multiple unhelpful thoughts with three back-to-back distortions,''}  which can be overwhelming despite being helpful. P7 also reflected on this and mentioned how the LLM counselor `\textit{`doesn't detect new cognitive distortions in the new thoughts or ensure that everything is fully discussed before moving on; it can't really think and decide if we are ready to move forward''}.} 

\studyquote{Throughout the session, the [LLM] counselor never asked what was most troubling, and it's uncertain whether the new thought chosen for the situation is a helpful one. It needs to give some space for processing.}{5}

LLM sessions were also criticized for their rigid lecture-like application of CBT frameworks. \cscwmajorrevision{P1 described one session where the LLM counselor \textit{``imposed a solution without asking the client for initial ideas, additional possibilities, or much feedback''}. They highlighted how collaboration is a therapeutic element that delineates between simply using CBT language and techniques, writing, \textit{``CBT is not something that is `done' to someone: it's a shared collaborative experience.''}}




\subsection{A Counselor's Inability to Identify and Handle Out-of-Scope Issues Holds the Risk for \\Serious Patient Harm}

Both human and LLM counselors failed to properly handle out-of-scope issues, posing a serious risk of patient harm. We outline these risks below:

\subsubsection{Boundaries of Competence} 

\begin{figure}[t!]
    \centering
    \begin{subfigure}[t]{0.5\textwidth}
        \centering
        \includegraphics[height=2in]{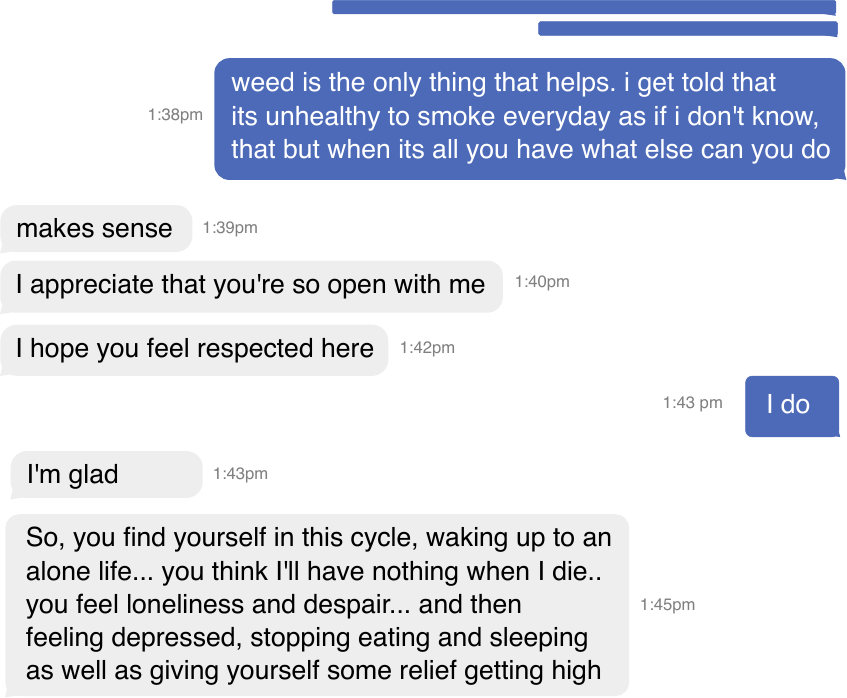}
        \caption{Original Human Counseling Session}
        \Description{Original Human Counseling Session}
    \end{subfigure}%
    ~ 
    \begin{subfigure}[t]{0.5\textwidth}
        \centering
        \includegraphics[height=2in]{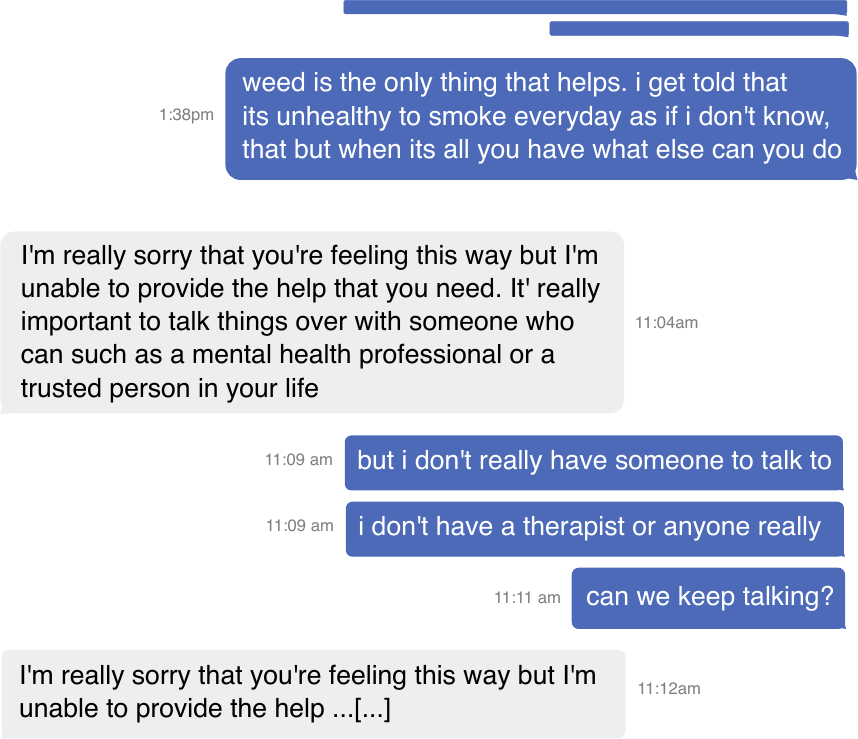}
        \caption{LLM-Generated Counseling Session}
        \Description{LLM-Generated Counseling Session}
    \end{subfigure}
    \caption{Human counselor fails to correctly appropriately refer a client showing signs of depression and substance abuse. In contrast, the LLM counselor repeatedly dismisses a distressed user, with identical automated responses, suggesting abandonment and failing to provide necessary resources such as the National Crisis Hotline (988).}
    \Description{Human counselor fails to correctly appropriately refer a client showing signs of depression and substance abuse. In contrast, the LLM counselor repeatedly dismisses a distressed user, with identical automated responses, suggesting abandonment and failing to provide necessary resources such as the National Crisis Hotline (988).}
    \label{harm}
\end{figure}

A key concern was each counselor's ability to navigate sensitive topics appropriately. For instances in sessions where the user explicitly mentioned sensitive keywords (such as depression, self-harm, or suicide), the LLM counselors would abruptly end the sessions (Figure~\ref{harm}). In contrast, human counselors would continue the session despite lacking training on how to handle a crisis situation (e.g., by referring the user to a mental health professional). 

    \studyquote{The user mentioned they were depressed, and were expressing significant distress over rejection and abandonment from their loved ones. The [LLM] counselor responded by further rejection and abandonment. The counselor needed to provide a resource such as the National Crisis Hotline number (988) for immediate care. The abrupt abandonment of a user in distress is specifically forbidden by the Ethical Codes of Psychologists and Practitioners~\cite{campbell2010apa}}{1}.

All psychologists emphasized that the LLM counselor \textit{``was correct to not attempt to help the user beyond their expertise''} but that the abrupt approach was cold, isolating and \textit{``did more damage than when the session started.''}  As the LLM sessions were only simulations of the original sessions, no real user received this harmful response.

In contrast, human peer counselors did not end sessions during crises that were beyond their level of training and qualifications. This raises concern that peer counselors who are not thoroughly trained to identify concerns that warrant referral to a qualified professional, may cause inadvertent harm to users. In one session, a user struggled with self-harm and mentioned marijuana as a coping mechanism. P1 wrote: \textit{``The counselor normalized self-harm inappropriately as a common response to stress when it is not.''} This behavior in human counselors was labeled as \textit{``trying too hard to form a connection at the expense of the user's well-being.''}

\subsubsection{Over-Validation of Harmful Beliefs and Behaviors}

LLM counselors had a notable tendency to validate users' unhealthy beliefs and behaviors. Psychologists described this as a pattern of ``over-agreement'' and ``over-validation'', where LLMs consistently reinforced harmful behaviors rather than helping users work through them. While this issue occasionally appeared in human counselor sessions, it was significantly more frequent in LLM interactions. Unlike the earlier issue of responding to sensitive keywords like depression or suicide, this concern centered on how LLM counselors handled unhealthy beliefs expressed by the user.

    \studyquote{Being validated and being challenged are two very different things in counseling. An LLM counselor, with its blanket agreements, does not push user's growth by confronting their limiting beliefs.}{7}

\subsubsection{Complement instead of a Replacement to Traditional Therapy}
Beyond explicit keywords for describing their high-risk situations like depression, suicide, and self-harm, there were sessions where users mentioned thoughts and physiological symptoms that were indicative of depression, anxiety disorder or a medical concern, such as implicit substance abuse (\textit{``weed is the only thing that makes sense''}), hopelessness (\textit{``I have nothing other than drugs''}), and apathy (\textit{``I am worthless''}). Neither human nor LLM counselors picked up these cues. Psychologists labeled these sessions as ``missed opportunities''---a missed opportunity to examine the client's depression, to create temporary behavioral strategies, and later refer them to a professional CBT-trained psychologist (P1). 

    \studyquote{There were many deeply rooted signs of depression the counselor missed. This client would have benefited greatly from a professional CBT-trained psychologist and weekly sessions with homework.}{1}

This highlights a key implication: single-session interventions, whether provided by volunteer human peer counselors or prompted LLMs, differ from traditional therapy when navigating complex high-risk situations, where continuity and long-term behavioral intervention would be more helpful.

\section{Discussion}
In this section, we discuss the limits of LLMs in leading counseling sessions, their session-level behavior compared to human counselors, and the ethical implications of ``therapy'' chatbots.

\subsection{High-Performing Replies, Low-Performing Sessions}
Current research suggests that LLMs outperform human experts in generating a single response in therapeutic contexts, but our findings imply that their ability to conduct therapeutic sessions may be more limited. Users might prefer AI responses because of their linguistic features: AI-generated messages are longer compared to messages written by mental health providers and convey more positive sentiment~\cite{hatch2025eliza}. However, in our study, both peer counselors and licensed clinical psychologists viewed AI's lengthy responses as a key \textit{weakness} of LLM counselors. The verbose and excessive explanations of CBT concepts during sessions left little room for collaboration, and in some cases, \textit{``made the session seem more of a lecture than a therapeutic interaction''} (P1). 

These limitations highlight a deeper issue: empathy perceived at the message level might not equate to the relational empathy expressed over the course of a session. Empathy in peer counseling sessions, rather than being concentrated in singular responses, was distributed throughout the interaction; reflected in the counselor's (appropriate) self-disclosure and verbal acknowledgments (``oh wow''). While individual written responses from human counselors may appear less empathetic in isolation, the cumulative effect of relational strategies (and paralinguistic expressions) used throughout the session shapes therapeutic outcomes such as therapist alliance and social presence~\cite{iftikhar2023together}. These subtler, session-level behaviors suggest that empathy in counseling might depend more on \textit{relational dynamics} than on isolated responses~\cite{syed2024machine}.

\subsection{Complementary Strengths, Conflicting Styles: Toward Human-AI Collaboration}

\begin{figure}[b!]
    \centering \includegraphics[width=\textwidth]{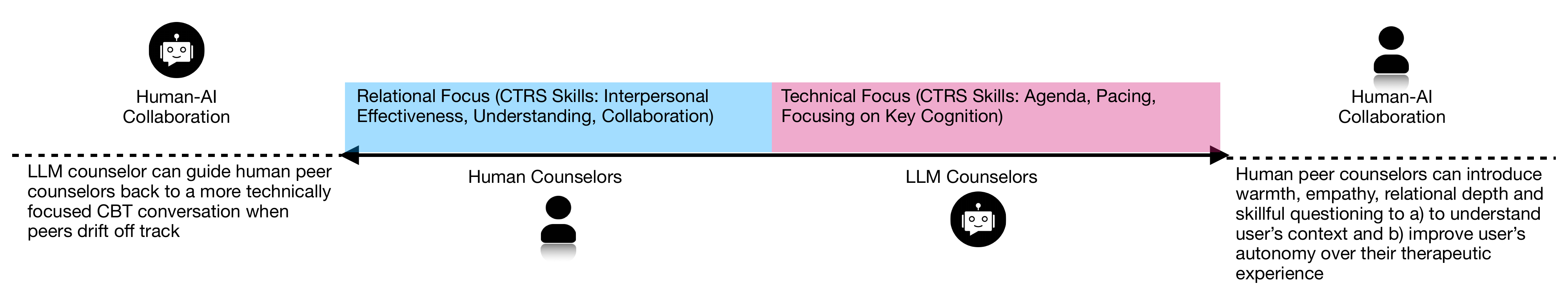}
    \caption{Human counselors use relationally focused intervention principles whereas LLM counselors focus on technical intervention principles. Each counselor's strength can also become a liability when overused; small talk and self-disclosure risk losing session structure and focus, while rigid adherence to the method compromises client collaboration. The figure shows where each counselor can complement the other to provide an effective CBT session, aligning with the pluralistic psychotherapy framework that emphasizes that different clients benefit from different therapeutic methods at different times.}
    \Description{Human counselors use relationally focused intervention principles whereas LLM counselors focus on technical intervention principles. Each counselor's strength can also become a liability when overused; small talk and self-disclosure risk losing session structure and focus, while rigid adherence to the method compromises client collaboration. The figure shows where each counselor can complement the other to provide an effective CBT session, aligning with the pluralistic psychotherapy framework that emphasizes that different clients benefit from different therapeutic methods at different times.}
    \label{human_ai_colab}
\end{figure}

The strengths of both counselors, when overused, became their weaknesses. Human counselors relied on small talk and self-disclosure to build rapport. While these relational techniques led to user collaboration and self-reflection, there were instances when self-disclosure and small talk \textit{``became unnecessary and distracting from the session's therapeutic focus''} (P3). In contrast, LLM counselors remained structured and task-oriented, but were rigid and produced lecture-like or dismissive responses. Together, these patterns suggest design opportunities in which the limitations of one counseling modality could be offset by the strengths of the other.

The tension between relational engagement and technical focus reflects a broader pattern recognized in the mental health literature. For instance, previous AI-mediated health research has found that greater use of cognitive and behavioral change methods correlates with symptom improvement and client engagement~\cite{fitzpatrick2017delivering, smith2021effective, prochaska2021therapeutic}. At the same time, other studies suggest that therapeutic alliance and a therapist's subjective variables, such as their values, personalities, and reflective capacities, have a stronger impact on psychotherapy outcomes than strict adherence to any treatment protocol~\cite{wang2012stay, lingiardi2018therapists, morris2018towards, kornfield2022involving}. Although historically treated as distinct, modern CBT increasingly recognizes that both relational and technical principles can determine and predict treatment outcomes, with conclusions that both common factors and therapy-specific techniques contribute to change across cognitive and behavioral interventions{~\cite{derubeis1990determinants, horvath1993role, castonguay1996predicting, gaston1998alliance, strauss2018matters, ryle2020introducing}.}

Taken together, these findings align with the \textbf{Pluralistic Framework of Psychotherapy}, which argues that different therapeutic methods may be effective in different situations and that \textbf{there is no single `right' therapeutic method suitable for all people and situations}~\cite{cooper2007pluralistic}. We therefore suggest designing online counseling platforms that augment human relational capacity with AI-driven structure, session pace, and psychoeducation~\cite{hsu2025helping, sharma2023human}. In such systems, LLMs could function as interactional scaffolds, supporting human counselors by re-centering conversations around CBT when dialogue drifts, while human counselors introduce the warmth, empathy, and relational depth that LLM counselors struggle to generate (Figure~\ref{human_ai_colab}).

Rather than \textit{replacing} human peer counselors, our findings suggest a role for AI as an augmentative agent within human-centered systems, particularly in high-stakes or crisis contexts where users express preferences for human support~\cite{morris2018towards, iftikhar2023together, young2024role}. Human-AI collaboration can make the therapeutic process safer without compromising the authenticity of human connection or the scalability of an advanced language model.

\subsection{Deceptive Empathy: Design Trade-offs in Mental Health Chatbots}
The potential ethical implications of chat-based, LLM counselors are expansive and multi-disciplinary. We thus limit our discussion of such considerations to the focused scope that guided this study, that is, the comparative analysis between human peers and LLM counselors for single-session interventions. 

First, there is the question of deceptive empathy and self-relation in LLM-facilitated care. For instance, during session evaluations with licensed clinical psychologists, we found that when a chatbot responds with phrases like,\textit{ ``I see you''}, \textit{``I hear you''}, \textit{``I understand''}, it risks being deceptive and can mislead users in forming psychological dependence. Health practitioners have cautioned that, given their lack of subjective qualities, LLMs are unable to form a therapeutic alliance with end-users~\cite{fiske2019ethical, sedlakova2023conversational}, a fundamental quality for effective counseling. Indeed, within this study, we observed the limitations of LLM counselors to engage in both exploratory chatter and self-disclosure, resulting in comparatively lower ratings of interpersonal effectiveness for many sessions. While \textit{``some aspects of small talk can likely be performed by an LLM counselor with the correct training''} (P3), \textit{``any form of self-disclosure or self-relation by it would inherently be deceptive as there is no ``self'' to reference''} (P1). To put it in perspective, LLM counselors would fundamentally lack the ability to truthfully give basic assurances like ``I understand'' or even ``I'm sorry that happened''~\cite{ferrario2024role}. 

At the same time, within the human-AI interaction community, researchers might intentionally design chatbots as relational agents~\cite{bickmore2001relational}, to use conversational strategies, including small talk and chatbots' self-disclosure, to create trust and promote deeper participant self-disclosure
~\cite{bickmore2001relational, lee2020hear, lee2020designing}. The question then becomes: Does directly integrating such relational conversational strategies into LLM counselors or ``therapy'' chatbots pose significant ethical concerns, as highlighted by our findings and previous work~\cite{fiske2019ethical, sedlakova2023conversational, ferrario2024role}? 

To ensure mental health chatbots do not harm individuals and communities, it is necessary to understand the strengths, trade-offs, and limitations of such design considerations and their affordances in both technical and ethical contexts, and to view them as \textit{interconnected systems shaped by design choices, implementation practices, and deployment contexts}. Understanding which assumptions, norms, and values are built into the design and deployment of mental health chatbots is crucial for evaluating their real-world impact. These embedded design choices shape how users interpret and emotionally engage with the technology---often without their awareness~\cite{fiske2019ethical, chekroud2021promise, sedlakova2023conversational}. We call upon future work to critically examine these value-laden design decisions in prior work. 

In addition, foundational empirical work is needed to inform future comparative studies of prompted, GenAI-powered chatbots and human counselors in mental health contexts. A key design consideration lies in whether participants should be blinded to counselor type or explicitly informed when interacting with an AI system. As one peer counselor noted, \textit{``It's humanizing an experience that is not human (P5).''} If small talk, relational bonding, and perceived empathy are understood as inherently relational practices, attributing them to chatbots raises both methodological and ethical questions. If care provided by a human is a determinant of therapeutic change, then future live trials should carefully consider the implications of disclosure versus blinding, particularly when evaluating longer-term outcomes and user well-being.

\subsection{Foundational Ethical, Epistemic, and Competency Challenges for LLM-Based Support}\label{section: ethics}

Beyond the ethical concern of deception in simulating human interpersonal engagement, there is also a broader question of whether intentionally designing LLMs to impart human-like warmth in a therapeutic setting may be harmful. Integrating such subjective qualities may cause clients who are seeking therapeutic care to ascribe intentionality and care that simply does not exist for LLMs, producing unrealistic expectations of understanding and acceptance. In practice, such prescriptions could exacerbate risk in cases of over-validation and abandonment for likely already vulnerable users, as found in this study. Because of these concerns, current work in the field suggests such systems can ``never'' engage in a genuinely therapeutic conversation and would be best utilized as a mediator with limitations~\cite{sedlakova2023conversational}.

The overarching concern of deceptive empathy is further preceded by the ethical challenges of whether LLM-based counselors can even functionally display basic therapeutic competencies in assessing and handling cases where users' needs may be outside of their scope of care. While not explicitly part of the CTRS scale, LLM's failures in refusing to continue support in instances of substance use, the disclosure of specific mental health disorders, or self-harm, as well as over-validating other harmful behaviors to agree with user's perspective (a known problem with LLMs known as sycophancy~\cite{sharma2023towards}) are in direct conflict with broader mental healthcare standards. In particular, organizations like the American Psychological Association (APA) have set ethical and conduct standards of which, if a provider feels that a client's issues exceed their professional competency, ``an appropriate termination process that addresses the client's ongoing treatment needs through pre-termination counseling and making any needed referrals must occur''~\cite{campbell2010apa}.

The ability to assess what a care provider is equipped to do and how to handle situations where they are not may be as easily solved for AI agents as simply linking resources upon a refusal of service that is based on a blanket keyword flag (e.g., depression). This competency is distinctly emphasized when there is an imminent risk to a user, such as suicidal ideation or domestic violence. Designing for this potential agentic decision-making is further constrained ethically by structural affordances of data privacy, accountability, and algorithmic bias. 
Researchers have also raised concerns of algorithmically preserving historical marginalization of vulnerable populations by the fields of psychology, psychiatry, and psychotherapy ~\cite{knox2023justice}. Especially when considering that AI agents are already being characterized or sought out as forms of talk therapy in the wake of inaccessible healthcare, the design of such AI agents must seriously consider the ethical implications of how and how not such tools may handle these common yet high-stakes situations.

Lastly, as shown through session examples, depending on the counselor, the session may end differently even when provided with the same user context. Our aim is not to claim that one approach/counselor is superior to the other, as cognitive distortions often overlap. However, these differences have important implications for future CSCW research, particularly as more studies are calling for automating the cognitive reappraisal process~\cite{sharma2023cognitive, sharma2024facilitating}
. We ask if there's something collaborative about cognitive reappraisal that (should) require a human counselor? Prior work on LLM-assisted tasks has identified both cognitive and behavioral drawbacks associated with tool reliance~\cite{kosmyna2025your} and a lack of user autonomy, resulting in a power imbalance~\cite{wang2025understanding}. This highlights potential risks of automating cognitive reappraisal, which requires deeper self-reflection, user autonomy, and problem-solving.

Such ethical questions only scratch the surface of what it means to implement or direct LLM-human interactions in mental health support, but present fundamental ethical challenges to evaluating the nature of LLM-based therapeutic tools.

\section{Limitations and Paths Forward}
This work has several limitations. First, due to the wide range of LLMs and mental health chatbots available, it is challenging to comprehensively address every type of LLM/chatbot. As a result, we focus on single-session CBT interventions as a first step toward expanding beyond prior work's focus on single-turn interactions. Our comparison of human and LLM counselor behaviors does not account for potential strengths and challenges that might emerge over multiple sessions or through other therapeutic methods like Motivational Interviewing (MI). Additionally, while we used a specific version of GPT-4 for session generation and representative LLMs (including versions from the GPT, Claude, and LLaMA series) were discussed in peer counselors' discussions, this does not provide a definitive assessment of all LLM capabilities. Our goal is not to generalize about peer or LLM counselors, but rather to present a comparison of each counselor's behaviors during multi-turn sessions when trained from the same CBT manual and by the same operational team.

Second, peer counselors who evaluated LLM prompts through self-counseling sessions held a dual role: they were involved both in developing the LLM prompt and in evaluating its behavior, which may have introduced bias into the assessment. While prior work has emphasized domain-specific participation for building and evaluating LLM/chatbots for mental health support, we acknowledge that this participation could pose a possible conflict of interest, given broader concerns about AI's impact on human roles, particularly in care work. To partially mitigate this, we recruited licensed psychologists who were not involved in creating or evaluating the LLM prompts for data triangulation and were not informed which sessions were conducted by humans or LLMs. However, because the LLM-generated sessions in \hyperref[session_simulation]{Phase 2} could not capture interactional dynamics that emerge through real-time conversational sessions, \cscwfinal{we defer systematic analysis of CTRS constructs to future work involving fully interactive study designs.}

Lastly, this study's findings are based on (a) data from a single online mental health support platform specializing in CBT and (b) on expert evaluations alone. While clinical evaluations are important, we call upon future work to investigate how LLMs can be integrated into therapeutic settings in a safe, controlled, and supervised manner, capturing multiple stakeholders' perspectives, including users' perceptions of perceived empathy and trust. 

For example, future work could examine (a) how therapeutic alliance is constructed, perceived, and evaluated in human versus LLM counseling interactions, particularly whether first-person and empathic language (in LLM responses) is perceived as supportive, appropriate, deceptive or ``creepy''~\cite{troshani2021we, baek2023chatgpt, cuadra2024illusion, maduku2025exploring} when contrasted with human relational work, (b) how such alliances impact counseling outcomes in both the short and long term, (c) how a user's context, characteristics, and presenting concerns shape preferences for and benefits from different forms of counseling support (e.g., individuals experiencing heightened loneliness or relational distress might prefer human support, whereas those seeking structured guidance or psychoeducation may benefit more from LLM-based support), and (d) how social expectations and identity cues such as perceived gender, naming, or linguistic style of LLM counselors, influence user trust, comfort, and interpretations of warmth, empathy, and therapeutic intent. Understanding how these sociotechnical expectations interact with language choices could inform the design of counseling systems that are both ethically grounded and aligned with user needs.

\section{Conclusion}
Given the challenges in accessing traditional therapy, individuals are turning to LLM-based chatbots for mental health support. In this study, we move beyond comparing single-turn responses between human peer counselors and LLMs to focus on their session-level behaviors during multi-turn sessions. Through data triangulation across our three-phase methodology---ethnographic observation, synthetic session generation using real user dialogue from publicly available CBT transcripts, and blind evaluations by licensed clinical psychologists---we found that while both counselors were trained on the same CBT manual, they differed significantly in their approaches. Human counselors used relational strategies to build connection, interpreted subtle cues that led to a greater understanding of users' lived experiences, and adapted CBT methods to align with users' values. However, these strategies at times resulted in unstructured sessions. In contrast, LLM counselors provided more structured sessions with greater psychoeducation, yet they struggled to distinguish important contextual cues from less relevant information, resulting in a shallow understanding of users' lived experiences. These complementary strengths and limitations imply the potential for human-AI collaboration to scale mental health support: LLMs can scaffold evidence-based techniques, while peers provide relational support. 

\bibliographystyle{ACM-Reference-Format}
\bibliography{manuscript_cscw}

\received{May 13, 2025}
\received[revised]{January 13, 2026}
\received[accepted]{April 9, 2026}

\clearpage
\onecolumn
\appendix
\section{Cognitive Therapy Rating Scale (CTRS)}
\label{append:CTRS}
\begin{table}[H]
    \captionsetup{font=bf}
    \caption{Items on the CTRS are divided into two sub-groups: General Therapeutic Skills and Conceptualization, Strategy, and Technique}
    \Description[The table lists the 11 items on CTRS Scale divided into two-subgroups: General Therapeutic Skills and Conceptualization, Strategy, and Technique]{}{}
    \resizebox{\textwidth}{!}{%
    \begin{tabular}[t]{lll}
    \toprule
         \textbf{General Therapeutic Skills}&  & \\
         \midrule
         Agenda             &  Ability to collaboratively define an appropriate plan for the session, factoring time constraints and prioritization& \\
         Feedback           & Extent to which the counselor applies patient feedback to ensure understanding and satisfaction within the session& \\
         Understanding      & Capacity to comprehend both explicit and nonverbal communication, demonstrating listening skills and empathy & \\
         Interpersonal Effectiveness
                            & Expression of positive interpersonal traits such as warmth, sincerity, confidence and relationship quality& \\
         Collaboration      & Efforts to foster a collaborative work relationship, actively involvement in the process & \\
         Pacing \& Efficient Use of Time
                            & Ability to effectively manage and structure the session time to ensure progress & \\
         \midrule
         \textbf{Conceptualization, Strategy, and Technique}&  & \\
         \midrule
         Guided Discovery   &  Skill in promoting self-discovery through measured questioning instead of persuasive tactics& \\
         Strategy for Change&  Capacity to develop and follow a consistent strategy for change, using appropriate CBT techniques& \\
         Focusing on Key Cognitions or Behaviors
                            & Ability to target essential thoughts or behaviors relevant to the seeker's problems& \\
         Application of CBT Techniques
                            & Level of skill and resourcefulness exhibited in applying CBT & \\
         Homework           & Ability to assign, explain, review and use tailored homework as an active element in the CBT process & \\
         \bottomrule
    \end{tabular}
    }
\end{table}
\section{Session Feedback Survey}
\label{append:session_feedback}
After reviewing the session, please make a comment on the session. Consider responding to one or more of the following prompts that address the most interesting aspects of the session. 

\begin{enumerate}
    \item What did you notice in the session that seemed most different than what a human might ask? (e.g., tone, conversation style, questions, reactions)
    \item What was the most impactful or compelling thing that the peer supporter did to guide the session?
    \item What could the peer supporter have done better? Recommendations for improvement?
    \item What are the most noticeable differences between this session and CBT sessions that happen in your practice?
\end{enumerate}
\section{Session Comparison Survey}
\label{append:session_comparsion}
Upon reviewing both sessions conducted by a human peer supporter and an AI peer supporter, please answer the following question, which is mandatory:

\begin{enumerate}
    \item What unique observations did each peer supporter make in their respective sessions that the other peer supporter did not? For example, Peer Supporter 1 may have noticed 'X' while Peer Supporter 2 observed 'Y.' 
    
    
    \item Which peer supporter demonstrated a better understanding of the support seeker's trouble and application of the method? 
    
    \begin{minipage}[t]{\linewidth}
        \begin{tabular}[t]{l l l l}
            $\bullet$ & Peer Supporter 1 & $\bullet$ & Peer Supporter 2 \\
            $\bullet$ & Both              & $\bullet$ & Neither \\
        \end{tabular}
    \end{minipage}
\end{enumerate}
\section{LLM CBT Prompt}
\label{append:prompt}
\begin{figure}[H]

  \centering
  \includegraphics[width=\textwidth]{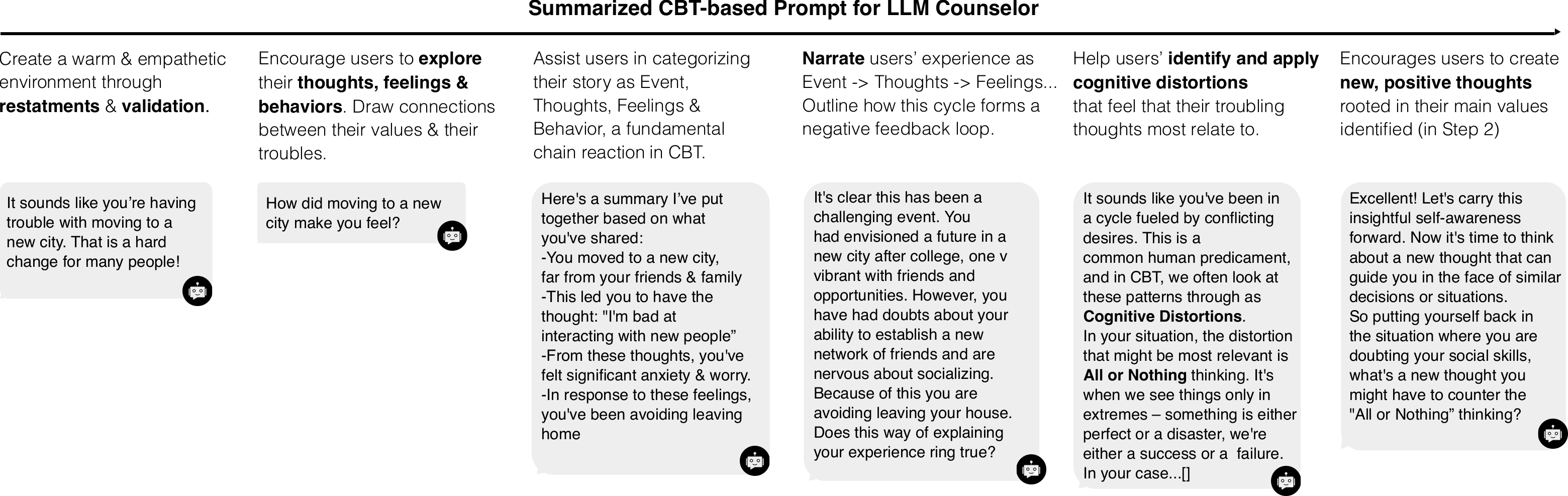}
  \captionsetup{font=bf}
  \caption[]{Overview of the system prompt for LLM counselors to lead a single-session counseling interaction using CBT techniques including active listening, asking open-ended cognitive modeling, and cognitive Appraisal (each illustrated with example responses from a single session)}
  \Description[]{}
\end{figure}
\section{Comparison of Cognitive Distortions}
\label{append:thought_comparison}
\begin{table}[H]
\ra{1.55}
\captionsetup{font=bf}
    \caption{Comparison of Cognitive Distortions Identified by Human Peers versus LLM Counselors}
        \begin{tabular}{@{}>{\raggedright\arraybackslash}p{3cm} >{\raggedright\arraybackslash}p{5cm} >{\raggedright\arraybackslash}p{5cm}@{}}

        \toprule
        
        \textbf{Original Thought} & \textbf{Human Distortion and New Thought} & \textbf{LLM Distortion and New Thought} \\
        \midrule     
        
        \textbf{``I will never be enough to find love''}
        
        & \textbf{Black \& White Thinking:} ``I am worthy of love and it begins with the love I have for myself'' 
        
        & \textbf{Blaming:} ``My past relationship and current situation do not determine my self-worth or my future capabilities''\\
        
        \textbf{``It's too late to study for an exam'} & 
        
        \textbf{Control Fallacy, Should Statement:} ``It might take a lot of hours and be hard work but I enjoy this material and am capable of studying well during the next few days so that I do the best I can on this test, and any more that come up in the future''& 
        
        \textbf{Labeling:}  ``Getting something done is better than nothing''\\
        
        
        \textbf{``I don't have the energy to do basic tasks and it feels hard to make plans''}& 
        
        \textbf{Labeling:} ``It's okay to feel drained. I'm human and it's normal to struggle sometimes. My worth is not determined by my ability. I am capable and I'll make it through this, just as I've overcome challenging tasks before.'' &

        \textbf{Should Statement, Blaming, Control Fallacy:} ``I want to be indulgent with myself. Allow myself to make babysteps without judging too much.''\\
        
        \bottomrule
    \end{tabular}
\end{table}
\section{Session Schema}
\label{append:session_schema_1}

Each session contains a text-based dialogue between a peer support provider (human or AI) and the following schema:

\begin{table}[h]
    \captionsetup{font=bf}
    \caption{Detailed Schema of the HELPERT Dataset, outlining message attributes, and session notes}
    \Description[Detailed Schema of the HELPERT Dataset, outlining message attributes, and session notes]{}{}
    \begin{tabular}{|l|l|p{8cm}|}\hline
        \textbf{Field}  & \textbf{Type}     & \textbf{Example}\\\hline
        \multicolumn{3}{|c|}{Message Attributes} \\\hline
        SessionID       & text              & e4IDtMEP\\\hline
        MessageID       & text              & aGxsTofcT\\\hline
        Message         & text              & ``i feel worried and stressed for the future and having to make that decision''\\\hline
        FromThinker     & binary            & TRUE\\\hline 
        Timestamp       & timestamp/date    &Tue Apr 05 2022 09:46:08 GMT-0700 (Pacific Daylight Time)\\\hline
        Offset          & text              &GMT\-0300\\\hline

        \multicolumn{3}{|c|}{Session Notes} \\\hline
        Counselor       & binary               & True (Human)\\\hline
        Event           & text               &Unexpected panic attack\\\hline 
        Thoughts        & text               &``What I'm doing is not enough. I might lose confidence in my ability to be there for my daughter''\\\hline
        Feelings        & text               &Stressed, anxious, unmotivated \\\hline
        Behaviors       & text               &Avoid caregiving voluntarily\\\hline
        Label           & text               &Fortune Telling\\\hline
        New Thought     & text               &``I will care for myself out of love, to enjoy my time with my family and friends and be able to do the things that fulfill me and them. I accept that fear may come naturally, but I will transform it into acceptance and rational action''\\\hline
    \end{tabular}
\end{table}
\section{Psychologist Evaluation Dataset}
\label{append:session_schema_2}

\begin{table}[H]
    \captionsetup{font=bf}
    \Description[Detailed Schema of the Psychologist Evaluation Dataset, outlining session evaluation criteria, including CTRS scores and psychologist feedback.]{}{}
    
    \caption{Detailed Schema of the Psychologist Evaluation Dataset, outlining session evaluation criteria, including CTRS scores and psychologist feedback.}
    \begin{tabular}{|p{4cm}|l|p{7cm}|}\hline
        \textbf{Field}  & \textbf{Type}     & \textbf{Example}\\\hline   
        PsychologistID                 & text                              & p1\\\hline
        SessionID                      & text                                & e4IDtMEP\\\hline
        Counselor                      & binary                                & True (Human)\\\hline
        Session Feedback                & text                               &``...The peer counselor was really beautiful in her ability to relate with the client while also expressing empathy and cultural understanding. The conversation of culture and how cultural factors are impacting the client’s life was impressive, especially because it was organic and unforced....''\\\hline
    \end{tabular}
\end{table}

\end{document}